\documentclass[aip,reprint]{revtex4-1}
\usepackage{latexsym}
\usepackage{color}
\usepackage{epsfig}
\usepackage{amsopn}
\usepackage{amsfonts}
\usepackage{amssymb}
\usepackage{amsmath}
\usepackage{subfigure}
\usepackage{physics}
\usepackage{braket}
\usepackage{epstopdf}
\usepackage{nicefrac}
\usepackage{siunitx}
\usepackage{textcomp}
\usepackage{etoolbox}
\usepackage{adjustbox} 
\usepackage[linkcolor=blue]{hyperref}
\usepackage{graphicx}
\usepackage{dcolumn}
\usepackage{bm}
\usepackage[utf8]{inputenc}
\usepackage[T1]{fontenc}
\usepackage{mathptmx}

\usepackage{multirow}
\usepackage{array}
\newcolumntype{R}[1]{>{\raggedright\let\newline\\\arraybackslash\hspace{0pt}}m{#1}}
\newcolumntype{C}[1]{>{\centering\let\newline\\\arraybackslash\hspace{0pt}}m{#1}}
\newcolumntype{L}[1]{>{\raggedleft\let\newline\\\arraybackslash\hspace{0pt}}m{#1}}
\usepackage{hhline}

\def\STIQPTwoCol{}

\begin{document}

\title{Shuttling-Based Trapped-Ion Quantum Information Processing} 

\author{V. Kaushal}
\affiliation{QUANTUM, Institut f\"ur Physik, Universit\"at Mainz, D-55128 Mainz, Germany}

\author{B. Lekitsch}
\email[]{b.lekitsch@uni-mainz.de}
\affiliation{QUANTUM, Institut f\"ur Physik, Universit\"at Mainz, D-55128 Mainz, Germany}

\author{A. Stahl}
\affiliation{QUANTUM, Institut f\"ur Physik, Universit\"at Mainz, D-55128 Mainz, Germany}

\author{J. Hilder}
\affiliation{QUANTUM, Institut f\"ur Physik, Universit\"at Mainz, D-55128 Mainz, Germany}

\author{D. Pijn}
\affiliation{QUANTUM, Institut f\"ur Physik, Universit\"at Mainz, D-55128 Mainz, Germany}

\author{C. Schmiegelow}
\affiliation{Departamento de Fisica, FCEyN, UBA and IFIBA, Conicet, Pabellon 1, Ciudad Universitaria, 1428 Buenos Aires, Argentina}

\author{A. Bermudez}
\affiliation{Departamento de Fisica Teorica, Universidad Complutense, 28040 Madrid, Spain}

\author{M. M\"uller}
\affiliation{Department of Physics, College of Science, Swansea University, Singleton Park, Swansea SA2 8PP, United Kingdom}

\author{F. Schmidt-Kaler}
\affiliation{QUANTUM, Institut f\"ur Physik, Universit\"at Mainz, D-55128 Mainz, Germany}

\author{U. Poschinger}
\affiliation{QUANTUM, Institut f\"ur Physik, Universit\"at Mainz, D-55128 Mainz, Germany}

\date{\today}

\begin{abstract}
Moving trapped-ion qubits in a microstructured array of radiofrequency traps offers a route towards realizing scalable quantum processing nodes. Establishing such nodes, providing sufficient functionality to represent a building block for emerging quantum technologies, e.g. a quantum computer or quantum repeater, remains a formidable technological challenge.  In this review, we present a holistic view on such an architecture, including the relevant components, their characterization and their impact on the overall system performance. We present a hardware architecture based on a uniform linear segmented multilayer trap, controlled by a custom-made fast multi-channel arbitrary waveform generator. The latter allows for conducting a set of different ion shuttling operations at sufficient speed and quality. We describe the relevant parameters and performance specifications for microstructured ion traps, waveform generators and additional circuitry, along with suitable measurement schemes to verify the system performance. Furthermore, a set of different basic shuttling operations for dynamic qubit register reconfiguration is described and characterized in detail. 
\end{abstract}


\maketitle 

\section{Introduction}\label{sec:intro}

The potentially disruptive applications of quantum information processing have stimulated an intensive effort towards realizing fault-tolerant large-scale quantum computers during the last two decades. Originally developed in academic research institutions, there is nowadays a variety of small-scale quantum processors that have served for proof-of-principle demonstrations\cite{Ladd2010}. In the last decade, we have witnessed an impressive progress in the performance of such small-scale quantum processing nodes. Currently, research efforts are partially shifting to world-leading technology companies. The most promising platforms are currently superconducting circuits and trapped atomic ions \cite{Linke2017}. In this work, we focus on trapped-ion quantum computing (QC) platforms, which have been of paramount importance in the first demonstrations of some of the quantum algorithms that can potentially harness the power of quantum computers \cite{Monz2016, Debnath2016, Kaufmann2017}.\\ 

Progress in trap technology allows for long and stable storage of long ion strings, and improved laser addressing optics allow for conducting gate operations across the entire register \cite{Debnath2016}. Fast entangling gates \cite{Schaefer2018} allow for suppressing undesired couplings \cite{Choi2014, Green2015}, which opens up prospects for a scalable architecture based on multiple static ion registers \cite{Ratcliffe2018}. The crucial remaining question is how to scale these devices towards large-scale quantum processors that can actually overcome the performance of their classical counterparts. Although noisy intermediate-scale quantum  (NISQ) computing devices have reached a maturity allowing for proof-of-principle demonstration of quantum supremacy \cite{Arute2019, Pednault2019, Preskill2018},  commercially useful applications will still require substantial research and engineering efforts. In particular, the inherent fragility of quantum states ultimately requires the incorporation of quantum error correction (QEC) \cite{Terhal2015}, where trapped-ion platforms have also been used to demonstrate various methodological building blocks \cite{Schindler2011, Nigg2014, Linke2017}. However, scaling such platforms to register sizes and circuit depths allowing for scientific or commercial applications remains a formidable challenge. \\

Trapped ion QC platforms face two major challenges when attempting to scale to a large number of qubits. On the one hand, the quantum information processing  requires a very well controlled and extremely low-noise environment. This leads to the requirement of shielding and filtering, while sufficient access for control is to be maintained.  Furthermore, scaling qubit registers is demanding as the number of required 'control lines', e.g. electrical wires, laser beams or microwave fields, scales unfavorably as compared to classical information processing devices.\\

For current trapped-ion based QC realizations, the qubit register is given by a linear Coulomb crystal of atomic ions confined in an ellipsoidal ponderomotive radio-frequency (rf) potential \cite{James1998, Leibfied2003rmp}. With typical ion spacings in the few-micrometer range, the ions can be individually addressed with tightly focused laser beams \cite{Naegerl1999}, which allows for driving single- or multi-qubit gate operations on specific subsets of qubits. 
Multi-qubit entangling gates are the most demanding operations, which can be generated via radiation-induced coupling between the qubit states (internal electronic states of the atomic ions) and the collective vibrational motion of the Coulomb crystal \cite{Cirac1995}. Impressive results have been achieved using linear qubit registers, including entanglement of 14-qubits \cite{Monz2011}, 20-qubit quantum simulations \cite{Friis2018}, multi-species entanglement \cite{Negnevitsky2018}, remote entanglement of ions in two setups \cite{Stephenson2019}, demonstration of Groovers algorithm \cite{Figgatt2017} and an 11-qubit trapped ion quantum computer \cite{Wright2019}.\\  


When increasing the number of ions in a linear Coulomb crystal, several challanges arise: Smaller inter-ion distances deteriorate laser addressing capabilities, an increasing number of motional modes leads to spectral crowding thwarting gate interactions, the ion crystal approaches the zig-zag structural instability \cite{Landa2013}, and the overall difficulty of maintaining a large crystal stable in the presence of background gas collisions and electric noise increases. \\

\begin{figure*}[!htb]
	\centering
	\includegraphics[width=0.7\textwidth,trim={0 6,5cm 0 0cm},clip]{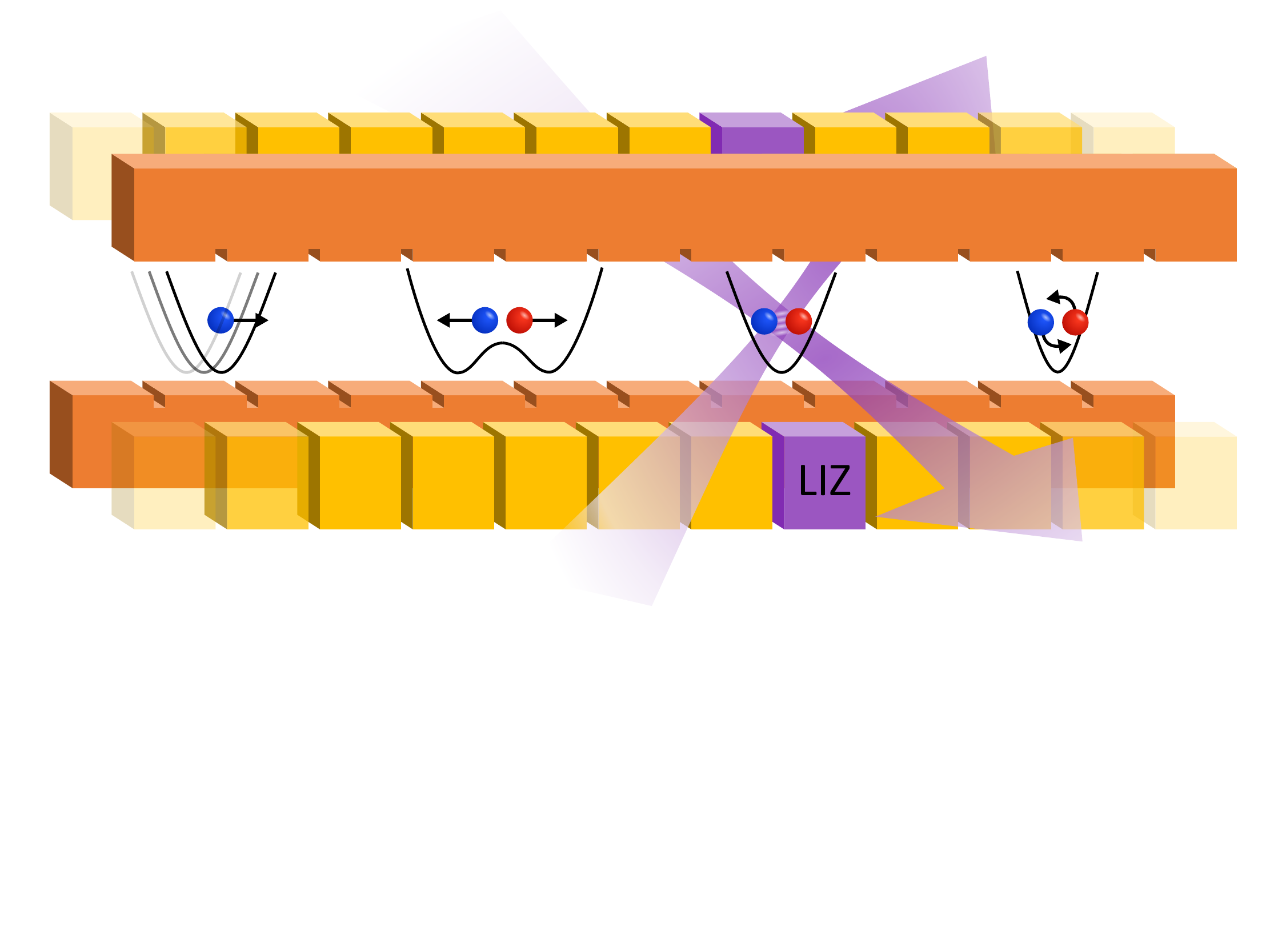}
	\caption{1D QCCD architecture: Trapped-ion qubits are arranged along the RF null axis of a uniformly segmented linear Paul trap. The lasers performing qubit  operations are directed to a fixed segment, the laser interaction zone (LIZ). The qubits are kept in small groups of two to four ions, and rearranged via shuttling operations, which are carried out by changing the voltages applied to the trap electrodes. The rearrangement operations fall into different categories, from left to right: Linear transport, separation and merging and swapping via crystal rotation.}
	\label{fig:architecture_overview}
\end{figure*}

These complications call for exploring alternative strategies to scale the trapped-ion quantum processor. To reach further scalability, the NIST group has proposed the \textit{Quantum CCD} as a tentative solution for a scalable trapped-ion QC architecture \cite{Kielpinski2002}: Here, a \textit{segmented} ion trap consists of an array of individually controllable electrodes. This allows for repositioning, commonly known as shuttling, of individual ions or small ion crystals between different regions of the trap, which can be processing zones allowing for qubit operations or mere storage regions. The degree of qubit connectivity can thus be adapted and optimized according to the requirements of the quantum algorithm to be executed. A recent theoretical analysis also shows that a quasi-1D qubit register topology with medium-range couplings, can be used to realize fault-tolerant operation at modest error-correction thresholds \cite{Li2018}. Such a quasi-1D register requires extensive use of shuttling operations.\\

Even with segmented traps, the achievable register sizes are subject to technological constraints. A possible route to further scalability is given by photonic interconnects, where light used to transfer quantum information between different quantum processing nodes \cite{Moehring2007, Monroe2014, Monroe2013, Brown2016, Pfister2016, Nigmatullin2016, Siverns2017}. Establishing such interconnects based on free-space or cavity-enhanced coupling is currently subject to intensive research efforts. Note that other schemes to scale trapped-ion QC architectures are currently under investigation. In particular, a complete blueprint for a shuttling-based two-dimensional array of surface electrodes traps has been proposed \cite{Lekitsch2017}, which is based on laser-less gates \cite{Weidt2016, Ospelkaus2011, Harty2016}. The basic functionality has already been demonstrated on a static three-ion crystal \cite{Piltz2016}.\\ 

In this work, we present a thorough description of a promising scaling strategy that can be considered a natural evolution of the static string-based approach, by exploiting shuttling and combining the 1D quantum CCD approach with optical interconnects for scaling. This scheme can alleviate the discussed limitations and problems of large static ion strings at the cost of increased technical complexity and overhead given by shuttling operations. We focus on a single trapped-ion quantum processing node, and provide a complete and detailed description of the technological developments that are required to realize a shuttling-based quantum information processing toolbox, which offers basic scalable functionality. We will partially review recent advances in the field of trapped ion quantum computing and provide references to seminal and recent work. Furthermore, our work contains a variety of recent and unpublished material serving as a summary of the current state-of-the-art concerning shuttling-based scalability and represents a road-map of technological advances required to further pursue this approach to scalable quantum information processing.\\

We start with a brief review of scalable trapped ion architectures with a focus on our approach in Sec. \ref{sec:architecture} and then describe the fabrication steps and properties of our linear segmented multilayer trap in Sec. \ref{sec:microtrap}, our custom-made multi-channel arbitrary waveform generator in Sec. \ref{sec:bertha} and the electrical wiring in Sec. \ref{sec:wiring}. The resulting system performance is presented in Sec. \ref{sec:characterization}. We then outline the methodology for conducting different types of shuttling operations in Sec. \ref{sec:shuttling}. Finally, in Sec. \ref{sec:outlook}, we give a conclusion and brief discussion on future improvements, enhancements and developments towards a scalable trapped-ion quantum processing node.

\section{Architecture}\label{sec:architecture}

All presented trapped-ion quantum computing architectures are based on radiofrequency (Paul) traps. Two variations of such traps are commonly employed \cite{Hughes2011}, one known as surface electrodes traps  \cite{Seidelin2006, Amini2010}, where all rf and dc electrodes are arranged in one plane below the trapping center, and a second configuration known as symmetric or sandwich traps, which has at least two planes with electrodes placed symmetrically above and below the trapping zone \cite{Stick2006, Blakestad2009}.

Architectures that allow scaling to large numbers of qubits without relying on optical interconnects make use of large electrode arrays connected using junctions \cite{Lekitsch2017, Kielpinski2002} and naturally favor surface trap designs. Proposals making use of optical interconnects \cite{Monroe2013} rely on smaller trap arrays and a smaller number of ions stored in each trap. Here, large arrays or junction designs are not ultimately required and therefore symmetric traps are favored due to superior ion confinement and easier trap operation.

The centerpiece of the architecture presented in this work is a symmetric two-layer segmented rf trap. The trap features two metalized electrode layers, each layer forming a linear array of uniformly sized dc electrodes arranged along the trap slit on one side and a single rf electrode placed on the other side. Each dc electrode can be individually supplied with computer-controlled arbitrary voltage waveforms. The direction along the trap slit is termed \textit{trap axis}, and the directions perpendicular to it are the \textit{transverse directions}. Two electrode layers, with the sides of dc and rf electrodes reversed, are placed symmetrically around the trap axis, one above and one below. An example of such a trap is shown in Fig. \ref{fig:architecture_overview}. In this design the confinement along the trap axis is of electrostatic nature, and can be controlled via the applied segment voltages. Application of a negative voltage to a given segment leads to an electrostatic potential well extending along the trap axis over the width of one segment, and a depth and potential curvature controlled by the voltage. The confinement along the transverse directions is of ponderomotive nature, and the axial confinement leads to weakening along at least one transverse direction, according to Laplace's law.

 Typical operation parameters are rf trap drive frequencies on the order of $2\pi\cdot\SI{35}{\mega\hertz}$, rf voltages of a about 200~V amplitude and dc trapping voltages around $-\SI{6}{\volt}$. Such parameters lead to axial secular frequencies of $2\pi\cdot\SI{1.5}{\mega\hertz}$ and transverse secular frequencies between $2\pi\cdot\SI{3.5}{\mega\hertz}$ and $2\pi\cdot\SI{4.5}{\mega\hertz}$. These confinement parameters allow for stable linear ion strings of up to four ions to be stored in one  segment potential. In total, we may operate on registers comprised of up to 20 ions.\\

Qubit register reconfigurations (i.e. shuttling operations) are carried out via temporal changes of the dc voltages, such that potential wells move along the trap axis. The fundamental challenge lies in realizing \textit{fast} operation times, in order to stay significantly below the decoherence timescales, and to keep the overhead as low as possible. Generally, shuttling operations do not have detrimental effects on qubit coherence, as recently shown by the Siegen group \cite{Kaufmann2018}. However, \textit{motional excitation} resulting from the shuttling operations is to be avoided, as it can deteriorate the fidelity of entangling gates or qubit readout operations. This directly implies a fundamental technological trade-off: On the one hand, fast operations require a large frequency bandwidth for the electrical waveforms to be applied to the dc electrodes, on the other hand, filtering of the dc lines is required to suppress technical noise. As a rule of thumb, we found that filtering at cut-off frequencies of 5-20\% of the lowest secular frequency to lead to suitable operating conditions.\\

In order to realize computational sequences with interleaved entangling gate and shuttling reconfiguration, we conduct the entangling gates such that they are mediated by and therefore only sensitive to the transverse motion of the ions. As the register reconfiguration operations predominantly lead to residual excitation along the axial direction, this scheme imposes much less stringent requirements on the reconfiguration operations. Note, that in actual experiments, we perform several hundreds of shuttling operations without the need for recooling the qubit register. Therefore, fast shuttling and bearable calibration effort can be successfully combined with entangling gate operations at fidelities in the $>$99\% regime.\\

All laser beams for conducting the various operations on the qubits are directed to a fixed segment, which is termed \textit{laser interaction zone} LIZ in the following. The computational sequences therefore consist of repeatedly regrouping the ions in a way such that gate operations or readout on a desired subset of the qubits can be carried out in the LIZ. We have demonstrated the basic functionality of the architecture by performing quantum-enhanced magnetic field sensing with separated, but entangled ions \cite{Ruster2017} and by generating a high-fidelity four-partite GHZ state \cite{Kaufmann2017}. So far, we have been working with two-ion crystals as the largest operational building block, however the extension to larger subsets is possible. Calculations indicate that the performance of a quasi-1D qubit register strongly increases with the maximum size of the subsets \cite{Li2018}.\\

The qubit type of our choice is the $^{40}$Ca$^+$ spin qubit \cite{Home2006}, where the quantum information is stored in the spin of the valence electron, i.e. the $\ket{S_{1/2},m_J=+\tfrac{1}{2}}$ and $\ket{S_{1/2},m_J=-\tfrac{1}{2}}$ sublevels of the electronic groundstate, which are split in energy by about $2\pi\times$~10~MHz by an external magnetic field. Our architecture is compatible with all trapped-ion qubit realizations, therefore this work does not elaborate on qubit-specific operations and restrictions. For more detail, the reader is referred to references \cite{Poschinger2009,Ruster2016}.
\section{Segmented microtrap fabrication}\label{sec:microtrap}
Ion traps can be manufactured using a variety of techniques, each having certain advantages depending on the dedicated purpose. For the storage of static linear strings of ions, large machined metal parts assembled to create a symmetric electrode configuration \cite{Schmidt-Kaler2003} still represent the most popular trap design. These traps are relatively easy to fabricate and operate. Shuttling ions in trap arrays requires a larger number of smaller electrode structures, leading to  microfabrication techniques being a more suitable choice.

Microfabricated ion traps can incorporate many trap features, including microwave lines close to the ion \cite{Ospelkaus2011, Allcock2013}, magnetic field gradient generating structures \cite{Kunert2014, Welzel2018}, micro-mirrors as part of the electrode surface \cite{Merrill2011}, on-chip detection \cite{Eltony2013}, complex electrode structures such as junctions\cite{Amini2010}, and other useful features \cite{Connell2017, Guise2015}.  Fabrication techniques for such traps range from CMOS (Complementary Metal-Oxide-Semiconductor) compatible methods \cite{Mehta2014} to standard MEMS (Micro-Electro-Mechanical Systems) microfabrication techniques \cite{Hughes2011, Seidelin2006, Amini2010, Wilpers2012, Clark2014, Ragg2019}. 

Micromachined symmetric traps, consisting of laser-machined and metal-coated alumina wafers, are designed and fabricated to accommodate many dc electrodes \cite{Blakestad2009, Turchette2000, Hensinger2006}, while some advantages of macroscopic linear rf traps retained, i.e. deep and symmetric ion confinement. Micromachined symmetric traps can be fabricated with minimal use of cleanroom fabrication techniques. The machining can be realized by employing commercially available laser cutting services, and near-perfect shielding of dielectrics can be achieved. 

\begin{figure}
	\centering
	\includegraphics[width=0.5\textwidth]{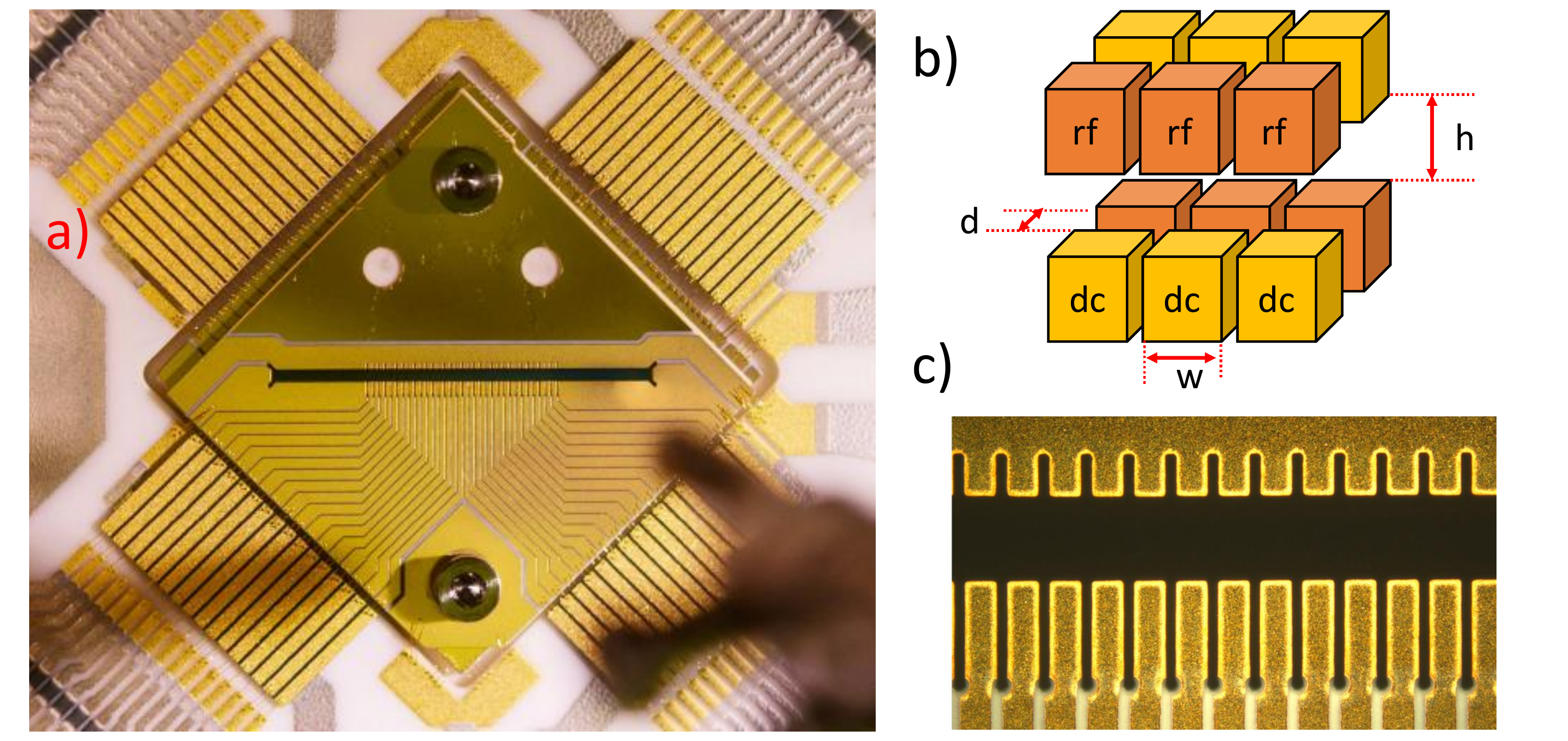}
	\caption{\textbf{a)} Close-up photograph of the trap after assembly and wirebonding. \textbf{b)} Sketch of the trap geometry, showing the relevant dimensions $w=\SI{150}{\micro\meter}$, $d=\SI{254}{\micro\meter}$ and $h=\SI{400}{\micro\meter}$. Note that these dimensions are reduced due to the electroplating of an $\SI{8}{\micro\meter}$ thick gold layer. The width of the isolation gaps between the dc electrodes is $\SI{50}{\micro\meter}$ before and on the order of $\SI{34}{\micro\meter}$ after plating. \textbf{c)} Microscope image of the trapping region. The exposed alumina between the electrodes serves as electrical isolation.}
	\label{fig:microtrap_schematic}
\end{figure}

For the experiments presented in this work a linear micromachined symmetric trap with 64 dc electrodes was used, which is depicted in Fig.~\ref{fig:microtrap_schematic} (a). The trap is made up of two electrode layers, which are stacked together using an additional spacer layer. The electrode configuration is shown in Fig.~\ref{fig:microtrap_schematic} (b). 
 
Fabrication of the electrode layers starts with polished alumina (Coors ADS996-polished / Al$_2$O$_3$ 99,6\%) as base material to achieve a low surface roughness. Individual parts are laser-cut from 2 inch alumina wafers forming the electrodes, trap slit and mounting holes. Cutting is performed using a high-power femto-second laser (Micreon GmbH, Hannover), which ablates material and provides clean cuts, allowing for structures down to the \SI{10}{\micro\meter} scale. After cutting, the wafers are cleaned using the following procedures: i) Ultrasonic cleaning in isopropyl alcohol at 40$^{\circ}$C for 20 minutes, ii) rinsing with distilled water, iii) submerging in piranha solution, 98\% H$_2$SO$_4$ to one part H$_2$O$_2$ (30\% dilution), at room temperature for 10 minutes, to remove organic and other reactive materials, iv) rinsing with distilled water followed by isopropyl alcohol and v) drying in air for 1 hour. 
   
Next, the chips are coated with a titanium adhesion layer of 50~nm, followed by a 500~nm layer of gold using an e-beam evaporator (Universit\"at Ulm, Institut f\"ur Optoelektronik) creating an oxide free and conducting surface for subsequent electroplating. To ensure homogeneous coating of all electrodes, the wafers are rotated during the coating process and the wafer normal is tilted by 20$^{\circ}$ with respect to the direction of incidence of the atomic beam.

In a subsequent laser cutting step at lower optical power, gold is removed from the alumina surface to create the electrode structures. In this step only the metals are ablated, while the alumina surface is retained. To ensure that no shorts appear at the intersection of electrode fingers and unstructured alumina, shown in Fig.~\ref{fig:microtrap_schematic} (c) an additional laser ablation step drills a hole at this intersection point, removing any metals deposited due to the angled evaporation and also alumina. At the top and left side of the trap, the electrodes remain fully connected along two edges of the trap, which is required for the following step. 

Our trap design and fabrication process is inspired by similar traps operated by the NIST group and the ETH Z\"urich group, where thick gold layers have been deposited via electroplating, and low heating rates were observed at room temperature. Thick electrode layers are desirable as the shielding of the trapped ions from the  dielectric wafers is improved, and the ohmic resistance of the dc and more importantly rf electrodes is reduced. Furthermore, the electroplated gold layer might exhibit more beneficial surface characteristics as compared to an evaporated layer. However, the dependence of heating rates on the trap surface characteristics is not yet fully understood. For instance, heating rates of below 10 quanta/s at secular frequencies in the 1~MHz range have also been observed in a trap with an evaporated surface of about 1~$\mu$m thickness by the Berkeley group \cite{Noel2019}.

\begin{figure*}
	\centering
	\includegraphics[width=0.7\textwidth]{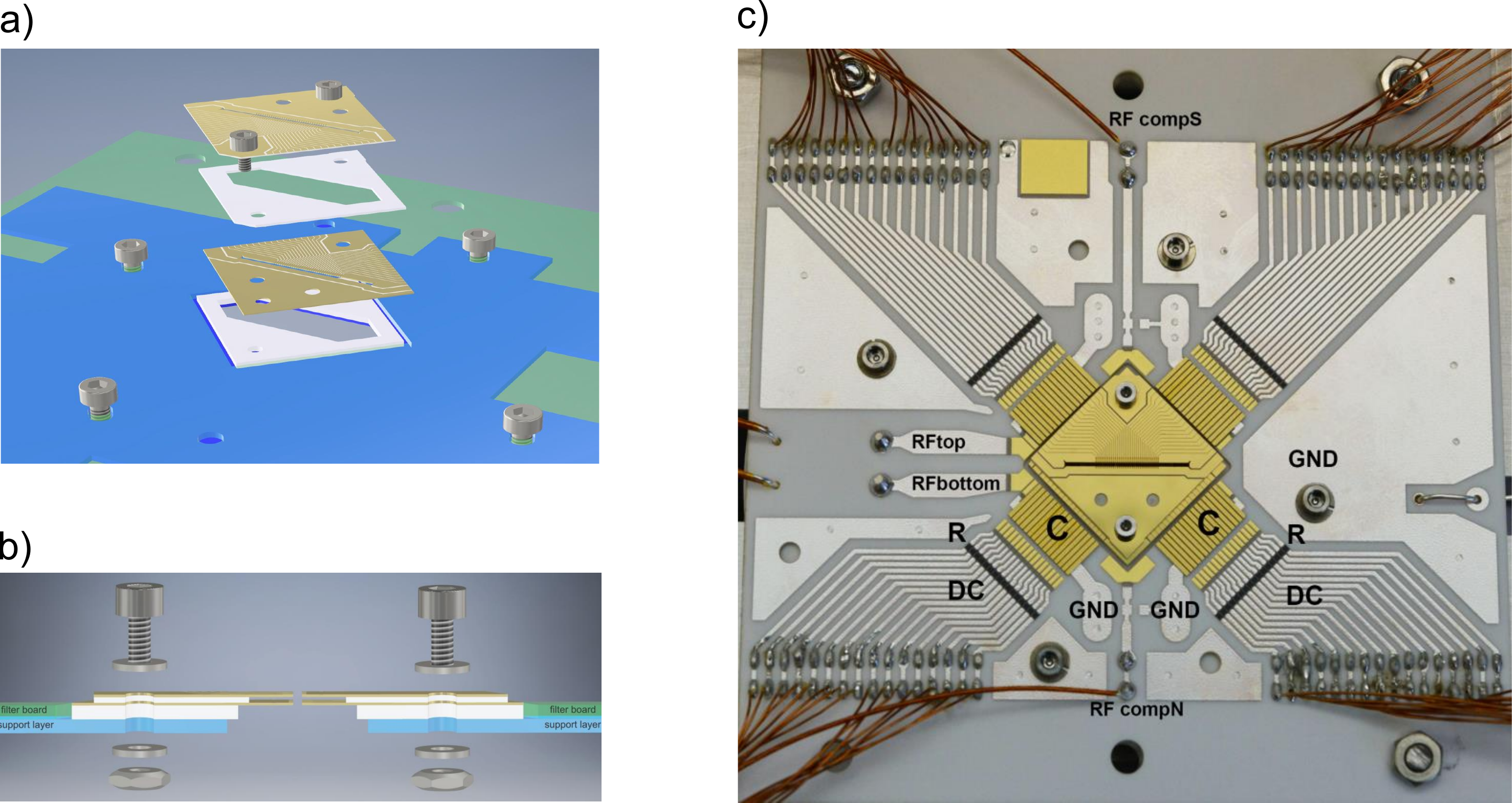}
	\caption{Trap assembly. \textbf{a)} and \textbf{b)} depict the assembly of the trap stack. The transparent green structure is the filter board, below which the support layer (blue) is mounted. Onto the latter, the following elements are mounted from bottom to top: Bottom spacer, metalized trap bottom layer, central spacer, metalized trap top layer. The stack is held together with M1 titanium screws. \textbf{(c)} shows a photograph of the complete electrical assembly.  The outer dimensions of the board are 79.9~mm x 78.3~mm. Each trap dc electrodes is connected to a single layer capacitor $C$ using wirebonds. Another set of wirebonds makes contact to printed metallized tracks followed by printed resistors $R$. Connections between the end of the tracks and vacuum feedthrough is made by soldering on Kapton insulated wires with a crimped on D-Sub connector.  The rf electrodes of the trap top and bottom layers are wired individually with cables of the the same length.}
	\label{fig:microtrap_trapassembly}
\end{figure*}

To achieve a thick gold layer covering the electrodes, the structured - but still connected - metal layers are gold electroplated to a thickness of about \SI{8}{\micro\meter}. The electroplating procedure is similar to the one established by the NIST group~\cite{Blakestad2010}. The solution used for the gold deposition is a sulphite-based gold bath Gold-SF (Metakem GmbH). Before the electroplating, the chips are cleaned in an ultrasonic bath of isopropyl alcohol (purity grade 99.8\%) at 40$^{\circ}$C for 15 minutes. We observed that if the chips are not cleaned within a few days after the last laser machining step, gold is more likely to be deposited on the electrically isolated parts during electroplating. This is likely caused by small particles of gold which are released during the laser ablation process and can then redeposit on gold-free areas if not cleaned in time. We also found that this problem can be avoided by submerging the traps in diluted iodine for a few seconds to dissolve these undesired gold particles without removing significant amounts of gold from the electrodes. After the electroplating, the trap chips are cleaned in distilled ionized water, isopropyl alcohol and acetone and inspected under a microscope. 

In the last step, chips are cut out using a saw blade to remove the electrical connections between electrodes, which were required during the plating process. Cutting these shorts was achieved using a 200~$\mu$m diameter diamond wire saw (Model 3500, Diamond WireTec GmbH). To perform the cutting process chips have to be glued onto the surface of a plastic mount using acetone soluble wax~(Crystalbond, Type 509). It is important that the trap surfaces are fully covered with the wax during cutting to prevent the wire saw from dragging gold across isolating parts of the alumina wafer which could otherwise electrically shorten the electrodes. Finally, the trap chips are removed from the plastic slab by dissolving the wax in acetone and then cleaned in an ultrasonic bath with distilled water followed by isopropyl alcohol.

After completing the individual trap chips, the symmetric ion trap is assembled using two trap chips and a mechanical spacer made of alumina. Traps chips and spacers are placed on a ceramic trap mount and aligned making use of an optical microscope, and fixated using two titanium M1 screws, as can be seen in Fig.~\ref{fig:microtrap_trapassembly}. This procedure achieves an alignment accuracy between the top and bottom trap chip of up to 10$\mu$m \cite{Huber2010}. Besides holding the trap in place and providing mounting structures to fixate the assembly inside a vacuum system, the ceramic mount is also used to establish electrical connections and provide low-pass filters.     

The trap mount is made of a polished alumina plate and has electrical tracks and resistors printed on it using conductive paste (made by Technische Universit\"at Dresden,  Fakult\"at Elektrotechnik und Informationstechnik), which is afterwards sintered, shown in Fig.~\ref{fig:microtrap_trapassembly} (c). To provide a low rf impedance path to ground for the dc electrodes and to create a simple low-pass rc filter, single layer ceramic capacitor arrays (Complex row capacitor, CR/CM series: CR16-200-344.2X178X10-3-G-202-Z) marked with C in Fig.~\ref{fig:microtrap_trapassembly} (c) are soldered onto the printed metal tracks. Both the top of the capacitor arrays and the electrical tracks of the trap chips are connected using wirebonds. Kapton insulated ultra-high vacuum compatible wires are soldered onto the end of the tracks on the trap mount and are connected to voltage waveform generators outside the vacuum apparatus, which will be described in the following chapters.

\section{Fast multi-channel arbitrary waveform generator}\label{sec:bertha}

Shuttling-based trapped-ion quantum processing requires freely programmable time-dependent voltage waveforms supplied to the individual trap electrodes. Therefore, a required key technological component is a fast multi-channel arbitrary waveform generator (mAWG), operating in real time, at sufficient channel count, memory depth and update rate. Moreover, stringent requirements are imposed on the electrical output characteristics.

Such mAWGs have been developed by several ion trapping groups and have been successfully employed to drive shuttling operations. These devices essentially consist of digital hardware receiving voltage waveform data from a control computer, storing these into a buffer memory and feeding them forward to digital-to-analog converters (DACs). The DACs and their subsequent output stages have the most significant impact on the performance of the system. Devices using parallel DACs have been developed by the NIST ion trapping and storage group \cite{Bowler2013} and the University of Siegen \cite{Baig2013}, other groups employ home-built \cite{Beev2017} or commercial systems using parallel or serial DACs.   

\begin{figure}
	\centering
	\includegraphics[width=0.45\textwidth]{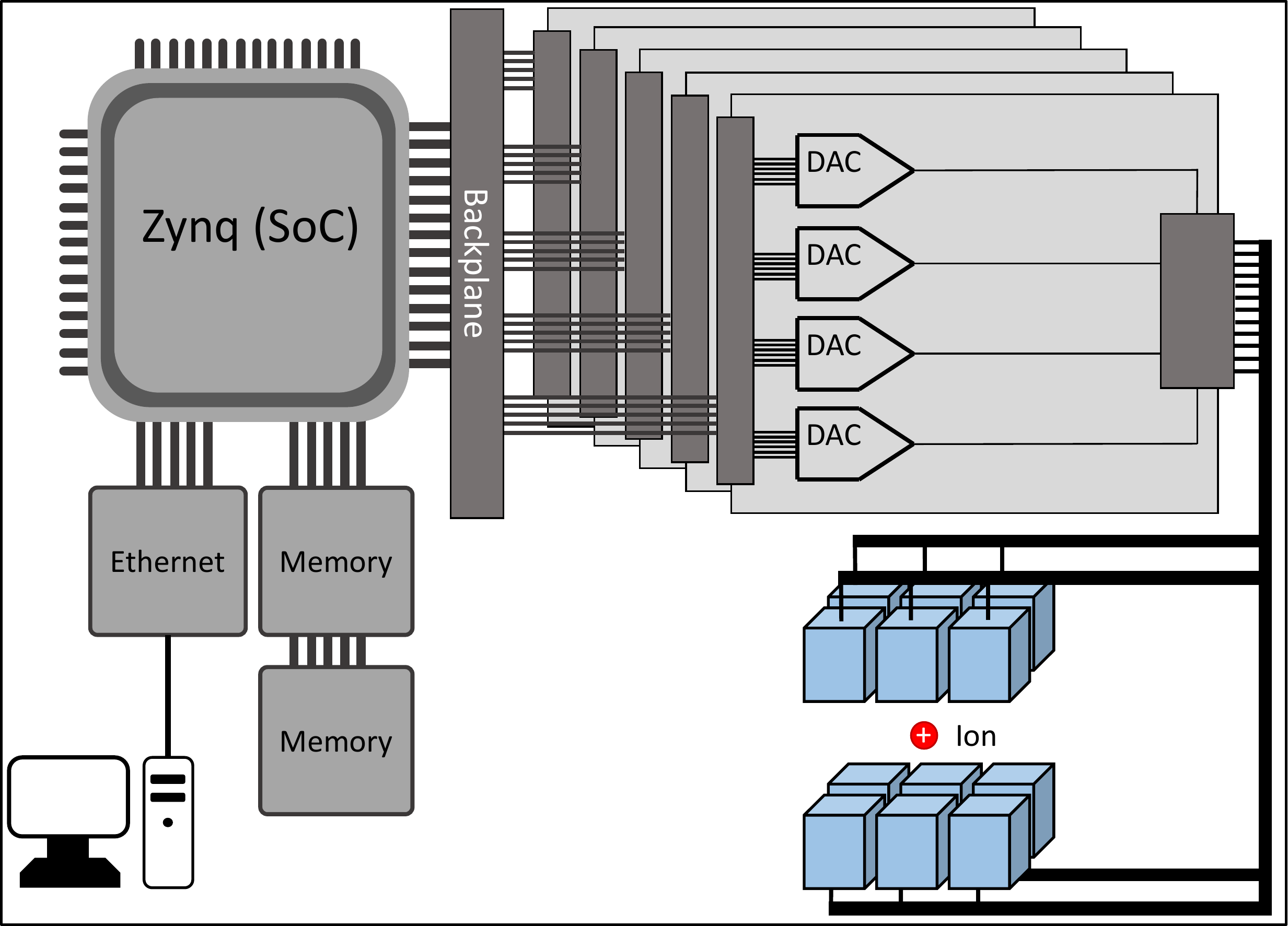}
	\caption{System overview showing connections between individual modules of the mAWG and also critical parts of the experiment, ranging from control computer to ion trap electrodes.}
	\label{fig:bertha_architecture}
\end{figure}

The mAWG developed in Mainz is described in this work and makes use of serial DACs. It is optimized for individual control of a large number of channels and employs a centralized \textit{Zynq} System-on-a-chip (SoC) device, which supplies digital waveform data to up to 80 DACs in real-time. It stores voltage waveform data on its on-board memory and supplies 24 additional digital I/O channels which can be used e.g. for laser control. A schematic showing the interplay of SoC, analog and digital modules and other essential components is presented in Fig. \ref{fig:bertha_architecture}. 

The mAWG is controlled and supplied with waveform data by a host computer using an Ethernet connection. The host computer sends data to CPU cores on the Zynq, which is stored in on-board dynamic random access memory (RAM) chips with a total storage capacity of 1 gigabyte. After receiving a trigger command from the host PC, the SoC streams the stored data to the connected DAC and digital output modules via the SoCs digital I/O pins. The DAC modules convert the digital signals to voltage waveforms. An overview of the general specifications achieved by the mAWG is given in Table \ref{tab:bertha_specifications}. More detailed specifications for the individual modules currently in use with the mAWG are given in the following sections. 

\begin{table}
\centering

\ifdefined\STIQPTwoCol
\begin{tabular}{ | L{5cm} | C{1.5cm } | C{1.5cm} |}
\else
\begin{tabular}{ | L{7cm} | C{2cm } | C{2cm} |}
\fi
    \hline
    \centering{\textbf{Characteristic}} & \textbf{Value} & \textbf{Unit} \\ \hhline{|=|=|=|}
    Max. data transfer rate & 730 & MBit/s \\ \hline
    Sequence memory & 900 & MB \\ \hline
    Max. internal clock rate & 400 & MHz  \\ \hline
    Signal clock rate & 50 & MHz  \\ \hline
    Sample hold time resolution & 20 & ns \\ \hline
   
    Min. analog sample update time & 360  & ns  \\ \hline
    Max. number of analog channels & 80 &  \\ \hline
    Output voltage range & $\pm$~40 & V  \\ \hline
    Min. analog resolution & 1.2 & mV  \\ \hline
    Slew rate & $\sim$ 20 & V/$\mu$s \\ \hline
    Noise floor & $<$100 & $\mu$V/$\sqrt{Hz} $\\ \hline
    \hline
\end{tabular}

\caption{Important performance specifications of the Mainz mAWG.}
\label{tab:bertha_specifications}
\end{table}

\subsection{Digital hardware}\label{sec:bertha_Zynq}

\begin{figure*}
	\centering
	\includegraphics[width=0.7\textwidth]{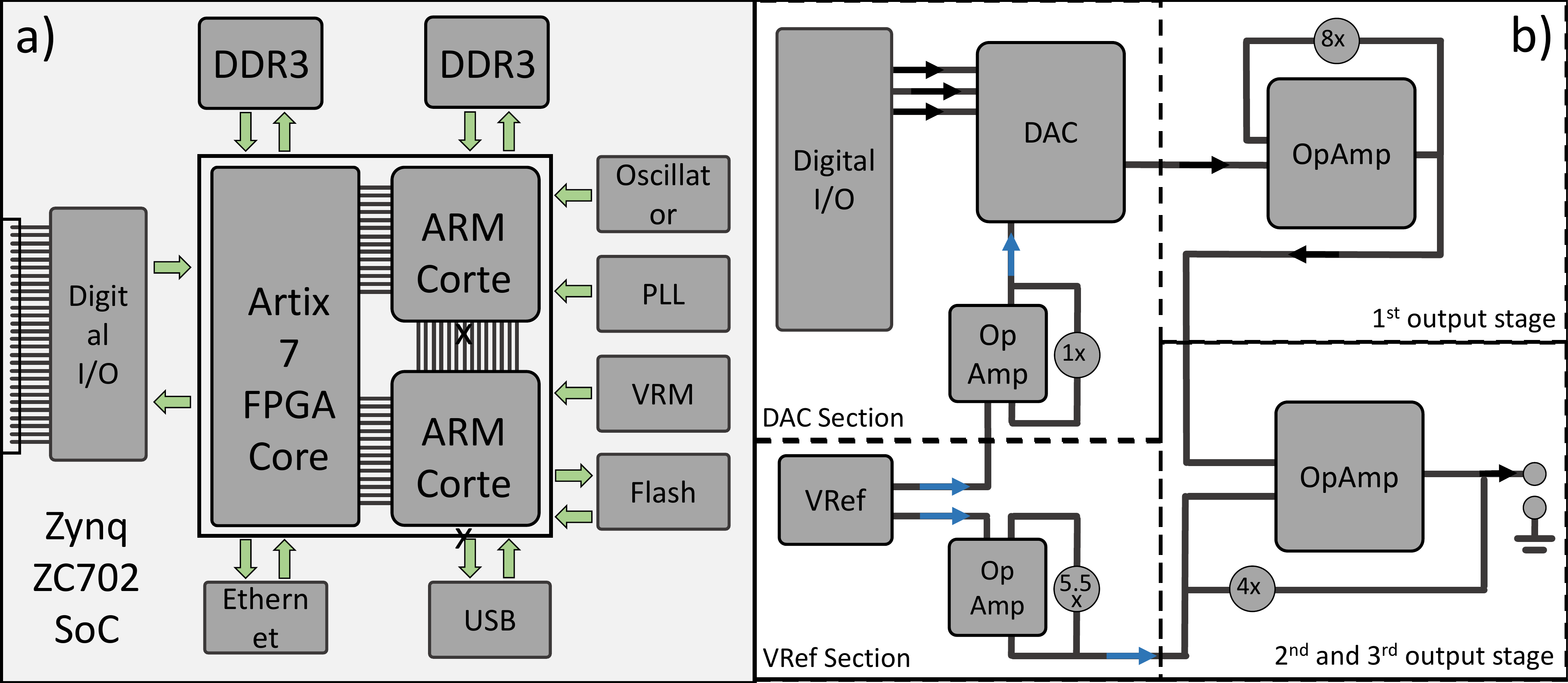}
	\caption{(a) Simplified overview of the Zynq ZC702 SoC evaluation kit showing the relevant on-board controllers and components. The connections between the SoC and external on-board components are illustrated. (b) Essential components of a single analog output channel. The DAC section features an AD5541 DAC supplied with data by the digital hardware and a reference voltage supplied by the VRef section, buffered by an OpAmp close to the DAC. The 1st output stage yields an 8x voltage amplification, followed by the 2nd and 3rd output stage which amplify the voltage further by a factor of 4 and perform a level shift.}
	\label{fig:bertha_analogue_Zynq}
\end{figure*}

We employ a Xilinx Zynq-7000 SoC ZC702 Evaluation Kit, which provides the Zynq SoC, memory and Ethernet interface, as the central control unit. It is based on the XC7Z020 Zynq chip, which combines a dual core ARM Cortex A9 667~MHz CPU and an Artix-7 FPGA with 85000 programmable logic cells. The SoC ZC702 board, illustrated in Fig.\ref{fig:bertha_analogue_Zynq} (a), features a total of 1~GB DDR3 memory, a 1~Gb/s Ethernet connector and corresponding controllers, a total of 200 digital in/outputs, on-board flash memory and other components. 

Communication between Zynq and host computer is established using an Ethernet-based TCP/IP connection utilizing the on-board Ethernet controller, managed by an ARM CPU core. The available USB connections are used to upload the drivers and software running on the ARM core and also to program the FPGA logic cells. During operation, data sequences are sent to the Zynq using the 1~Gb/s Ethernet connection, which is optimized for low latency ($<$1ms) and high throughput ($>$750 Mb/s). This ensures that experiments requiring long data sequences, exceeding the 1~GB on-board memory, can be run in an efficient way. 

On the host PC, a custom control software interfaces to all connected devices, including the mAWG. The control software converts voltage sample sequences into a binary format and sends them to the mAWG, which outputs these using a buffer system from the memory to the connected modules. The binary data is converted by the DACs to produce analog voltages. Additional execution control commands allow for interruption of the output by a predetermined duration, looping of a sequence and wait-for-trigger events. 

Harnessing the versatility of the Zynq, future developments will focus on implementing a more advanced storage and control system for the experimental sequences. These will be split into sub-sequences, and the selection and streaming will be initialized by an execution control logic running on the SoC. This will make it possible to perform autonomous branching of sequences, conditioned on measurement results processed by the mAWG. Additionally, the mAWG will be interfaced with rf waveform generators for laser pulse control, therefore realizing an autonomous real-time system capable of shuttling and gate operations conditioned on in-sequence measurement results - a cornerstone for quantum error correction protocols running on a scalable QC platform.

\subsection{Digital signal routing}\label{sec:bertha_Backplane and Digital Output Modules}

Routing of the binary data buffered on the Zynq is handled by a custom-developed back-plane. The Zynq I/O connections are directly connected to the back-plane, which distributes these to various digital I/O and analog output modules. 

For the signals supplying the digital I/O modules, the signal conditioning takes place directly on the back-plane. Digital I/O modules are connected to the I/Os of the SoC via the back-plane and to the front-panel outputs via level translation buffers (TI SN74LVC06A and SPC BSX20). The I/Os of the digital modules are used to control laser pulses and control of other experimental components. Each module features 20 digital outputs and four inputs with a maximum update rate of 50~MHz.

Analog output modules do not feature processing or conditioning of the digital input signals, therefore the back-plane is engineered to process digital and clock signals to a format usable by the DACs. To that end, the data and clock signal timings are adjusted to ensure that the stringent timing requirements of the DACs are met. Buffering clock drivers (TI CDC328AD) are used to distribute the main clock to all connected analog modules and their DACs. The issue of clock timing errors is addressed using a clock delay circuit (using a variable capacitor and ON 74ACT74D flip-flop), which can be manually adjusted to achieve the required clock timings with respect to the digital data. The digital signals are also buffered (ON 74ACT541SC-ND), which helps to correct for timing mismatches and maintains the required logic voltage levels. 

In addition, the DACs require a chip select signal, which initializes the shift of received binary data to the output stages. To reduce the amount of required digital I/Os, four DACs per module are supplied using a common chip select signal, which is buffered and distributed (ON 74ACT541SC-ND). While each DAC of a group can be updated independently, the update timings have to be the same, and also each DAC of a given group has to be updated even if the same voltage is to be maintained. The chip select, data and clock signals run through digital isolators (BCM HCPL-900J) in order to separate the grounds of the Zynq I/Os and the DACs. Galvanic isolation helps to prevent ground loops and reduces noise leakage from the digital signal processing and Zynq to the analog output modules.

\subsection{Analog output modules}\label{sec:bertha_mAWG Analogue Output Modules}

Analog output modules are equipped with DACs followed by output stages. Design choices for the used components are based on the requirements of performing fast shuttling operations: Voltage update rate of $\geq$ 2 million samples per second (MSPS), voltage range of $\pm$ 40~V, a resolution of 16~bit, stability of at least one least significant bit (80~V/ 16~bit), sufficient bandwidth and slew rate ($\geq$20~V / $\mu$s) at low noise levels. Fig. \ref{fig:bertha_analogue_Zynq} (b) gives a schematic overview of the individual components used for the analog output modules.  

For the most crucial component - the DACs - Analog Devices model \textit{AD5541} was chosen based on its update rate of 2.5 MSPS using a serial data interface, its low output noise of 11 nV/$\sqrt{Hz}$ and 16~bit resolution, with a relative accuracy of $\pm$~0.5 least significant bits (LSB). Significantly faster DACs are available, however these require a parallel data interface with 16 times more data lines per DAC. Other mAWGs used for control of trapped ions \cite{Baig2013, Bowler2013} feature parallel DACs, allowing for higher update rates, but require multiple FPGA controllers. For the intended operations being performed with this mAWG, the 2.5~MSPS update rate of the \textit{AD5541} is sufficient.

Each analog output module is equipped with 16 \textit{AD5541} DACs and receives 16 identical clock signals, 16 independent serial data and four chip select lines from the back-plane. The output range of the DACs is set to 0 - 2.5~V by a voltage reference (ADR441). Output voltages of the DACs require further amplification and the unipolar range needs to be converted to a bipolar $\pm$40~V range. In addition, a low output impedance is required to guarantee a sufficient bandwidth. 

The output circuit consist of two stages: 
\begin{enumerate}
\item
Pre-amplification with a gain factor of 8: An AD8510 op-amp and a resistor array are employed to achieve a stable gain factor, while ensuring that sufficiently low current is drawn from the DAC outputs. Despite the high gain of 8, this amplification stage maintains a large bandwidth of more than 3~MHz, while also reaching a slew rate of $\geq$20~V$/\mu s$. The AD8510 op-amps add further 8~nV$/\sqrt{Hz}$ noise to the signal generated by the DAC. The total noise, which will also include noise from clock signal bleed-through and Johnson noise from the used resistors will then be amplified by a factor of 8, putting the lower noise limit of the signal to $\sim$160 nV$/\sqrt{Hz}$ at this stage. 
\item
Level shifting and final voltage amplification with a gain factor of 4: The second stage uses an op-amp LTC6090, which features a maximum voltage output range of up to $\pm$ 70~V. Besides amplifying the input signal further by a factor of 4, a level shift from 0-20~V to $\pm$ 40~V is performed. This is accomplished using the on-board voltage reference, ensuring a symmetric bipolar output. The slew rate of $\sim$20~V$/\mu s$ at the chosen gain sets the limit of a full voltage swing ($+$40~V $\leftrightarrow$ $-$40~V) time to about 4~$\mu s$. Input voltage noise of the LTC6090 is specified to be 14~nV$/\sqrt{Hz}$, which adds to the minimum noise of  $\sim$160 nV$/\sqrt{Hz}$ on the signal after the first amplification stage. Total noise after the entire amplification stage will be characterized in more detail in the following section, but a lower limit is set by the $\sim$160 + 14 nV$/\sqrt{Hz}$ input noise, which is multiplied by the gain (4) to $\sim$700 nV$/\sqrt{Hz}$. 
\end{enumerate}
Each analog output module features 16 output channels accessible via a D-sub 37 pin connector.

\subsection{Noise characteristics}\label{sec:bertha_mAWG Noise Spektrum}

In this section, we present a detailed analysis of the analog voltage output performance of the mAWG, i.e. the noise characteristics in the frequency ranges particularly relevant for trapped ions, the long term voltage stability of the system, the inherent glitches and full voltage range pulse response. 

Noise spectra were recorded using an Anritsu MS2781B spectrum analyzer in combination with an active probe Agilent 41800A. Displayed minimum noise of the spectrum analyzer is -167~dBm at 0.1 Hz RBW.  The active probe (3~pF input capacitance and 100~k$\Omega$ input resistance) ensures that the measurement instrument load does not significantly affect the measurement results. Its frequency response was measured flat across the range 5~Hz to 500~MHz. 

\begin{figure*}
	\centering
	\includegraphics[width=0.8\textwidth]{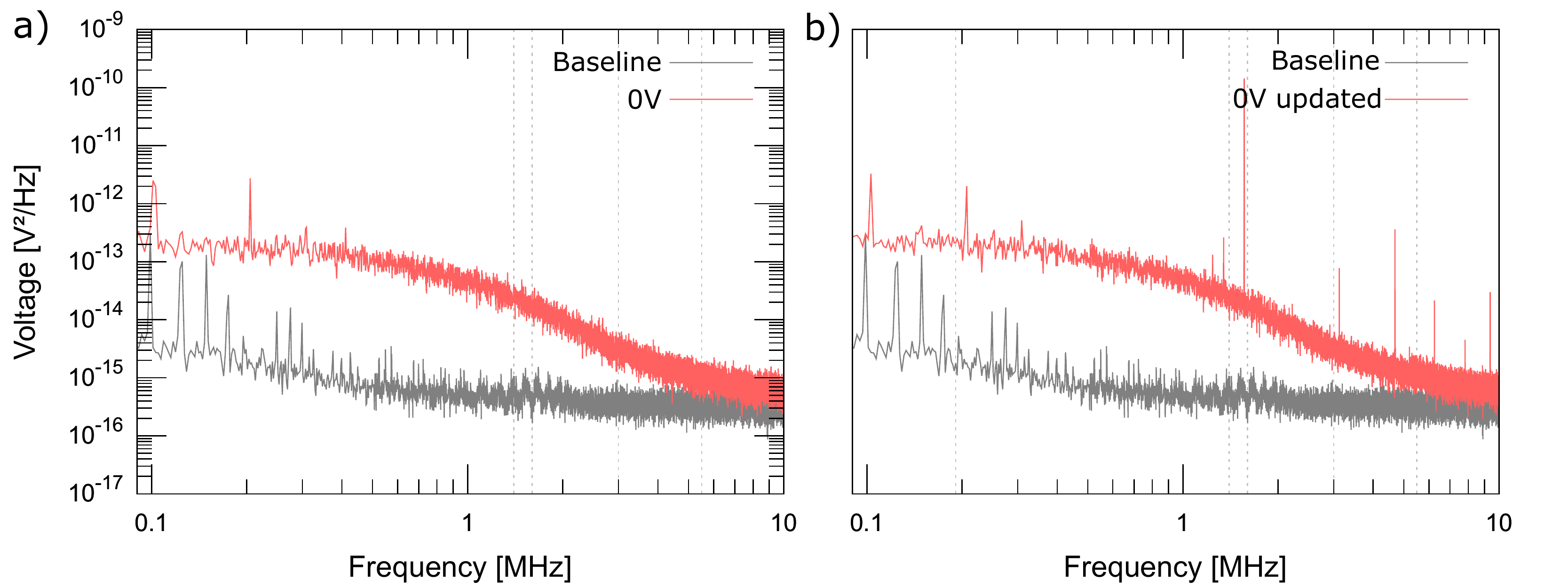}
	\caption{Spectra showing a baseline measurement and noise measurement for 0~V output without updates (a) and (b) with updates. The spectra display the expected $1/f$ dependence with additional pickup signals, originating from switch mode power supplies (100~kHz and 200~kHz). The spectrum with updates (b) also shows the update frequency close to 1.7~MHz and its higher harmonics. The vertical dashed lines indicate typical secular trap frequencies at about $2\pi\times$~1.5~MHz,  $2\pi\times$~3.5~MHz and  $2\pi\times$~4.5~MHz.}
	\label{fig:bertha_noise}
\end{figure*}

To establish a reference noise spectrum, the input noise floor was recorded with the bare active probe connected to the spectrum analyzer input. Then, spectra were recorded for 0~V voltage output from the mAWG with and without sending (constant) digital updates, as shown in Fig. \ref{fig:bertha_noise} (a). At 0~V the DACs and first amplification stage are at half scale output and are not expected to perform better or worse compared to other output voltages. Generating a static voltage without digital updates, see Fig. \ref{fig:bertha_noise} (a), represents an ideal case for low noise performance corresponding to the scenario when ions are statically confined in the trap. Outputting static voltages with constant digital updates of the same digital data is a common case when some voltages are kept constant and others need to be switched for shuttling, see Fig. \ref{fig:bertha_noise} (b). Digital updates need to be sent even if the voltage is kept static, since multiple DACs share a chip select channel. A noise spectrum for this case is presented in Fig. \ref{fig:bertha_noise} (b).  

The measured noise exhibits a significant increase towards lower frequencies, which is expected due to flicker noise, following a 1/f behavior, combined with Johnson noise stemming from the input resistances. Johnson noise is frequency independent, but a 1/f behavior is expected taking into account the passive RC load given by the active probe. Furthermore, the noise peaks visible at 100~kHz and 200~kHz can be explained by the presence of active switching power supplies in the vicinity of the measurement setup. 

The output noise spectra shown in Fig. \ref{fig:bertha_noise} display a noise floor on the order of $\sim$3 $\mu$V$/\sqrt{Hz}$ in the frequency range below 1~MHz, which is significantly higher than the noise floor of the measurement equipment. In the frequency range around 1~MHz the noise converges towards the level of the equipment noise floor and becomes indistinguishable at about 10~MHz.

The DACs in combination with op-amps are expected to contribute at least $\sim$700 nV$/\sqrt{Hz}$ of noise within the bandwidth of the amplification stages (2~MHz). At low frequencies (1~kHz) the reference voltage circuit will increase the minimum noise floor to $\gtrsim$~2~$\mu$V$/\sqrt{Hz}$.

The observed voltage noise spectral densities $S_V(\omega_j)$ would result in motional heating rates $\Gamma_h$ for a given mode $j={x,y,z}$ at secular frequency $\omega_{\xi}$ in quanta per second via the relation \cite{Brownnutt2015}

\begin{equation}
\Gamma_h=\frac{e^2}{4m\hbar\omega_j} \xi_j^2 S_V(\omega_j).
\end{equation}

Here, the $\xi_j\approx 1000\dots 2000$~m$^{-1}$ are geometric factors characterizing the electric field generated by the neighboring trap electrodes at the ion location, pointing along direction $j$.  We therefore estimate axial heating rates of up to 10$^4$ quanta per second, which is much higher than acceptable. We employ additional passive filters to reduce the voltage noise spectral densities by several orders of magnitude and will be discussed in chapter \ref{sec:wiring}. 

When digital updates corresponding to the same voltage are sent to the DAC, the noise, depicted in  Fig. \ref{fig:bertha_noise} (b), shows a clear increase  to $\sim$10~$\mu$V$/\sqrt{Hz}$ at the update frequency of $\sim$1.7~MHz and its higher harmonics. By appropriate choice of the secular frequencies, update-noise induced excitation can be avoided. \\

\begin{figure}
	\centering
	\includegraphics[width=0.4\textwidth]{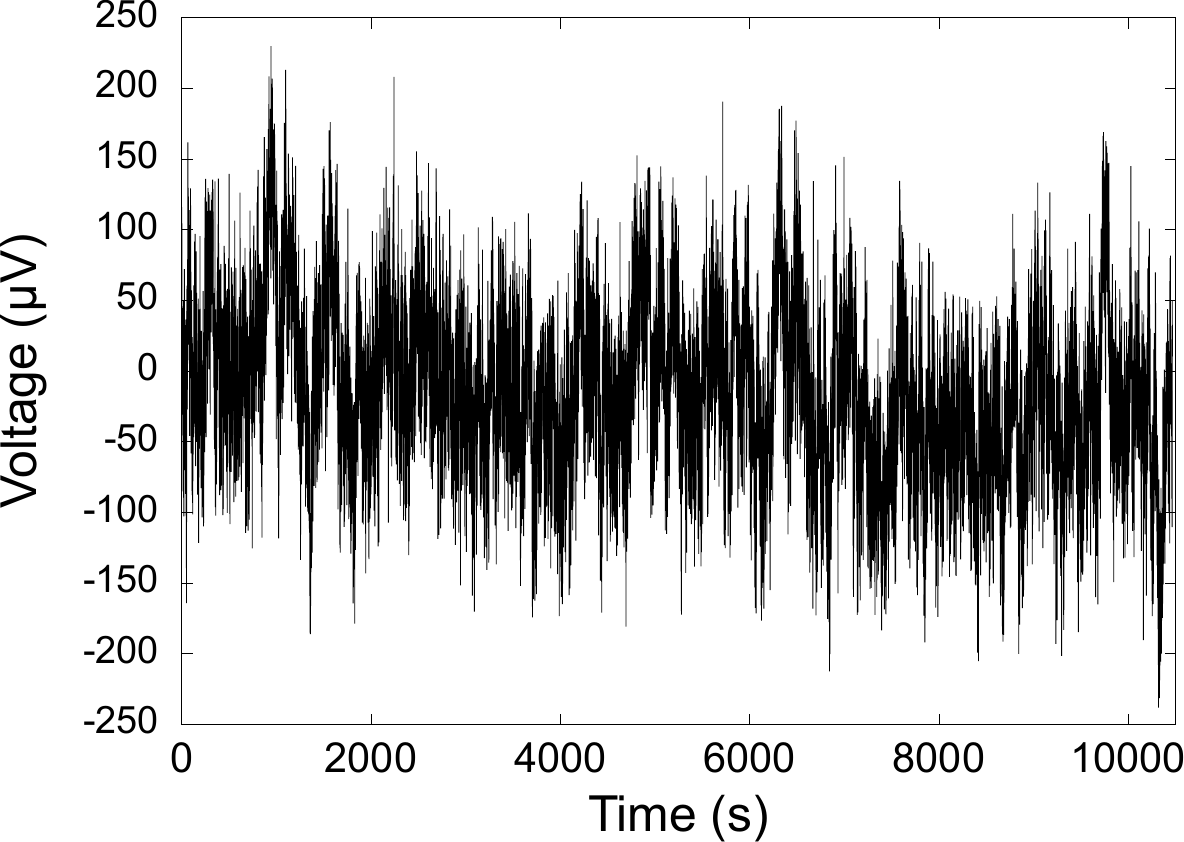}
	\caption{Voltage stability recording showing slow and fast voltage fluctuations measured each second over a time period of 10000 seconds. Slow drifts, even over this large time period, are significantly smaller ($\sim$50~$\mu$V) than faster drifts ($\pm$ 200~$\mu$V).}
	\label{fig:/bertha_voltagestability}
\end{figure}

Long term voltage stability and output voltage accuracy were characterized using an Agilent 34411A 6.5 digital multimeter. The output voltage was set to 0~V in order to to use the smallest measurement range $\pm\SI{100}{\milli\volt}$ with the lowest measurement noise floor of  3~nV and best measurement accuracy of  100~nV, which is well below the LSB of the mAWG of $\sim\SI{1.2}{\milli\volt}$. Voltage samples were recorded over a period of more than 160~min at a sampling rate of 1/s, the result is shown in Fig. \ref{fig:/bertha_voltagestability}. The rms voltage deviation of about $\SI{140}{\micro\volt}$ on timescales beyond 1~s is smaller by a factor of 10 compared to 1~LSB of the mAWG. Voltage drifts will detrimentally affect entangling gates due to a mismatch between the preset and actual secular frequencies \cite{Ballance2014} and motional dephasing, see Sec. \ref{sec:motionaldephasing}.

\begin{figure*}
	\centering
	\includegraphics[width=0.7\textwidth]{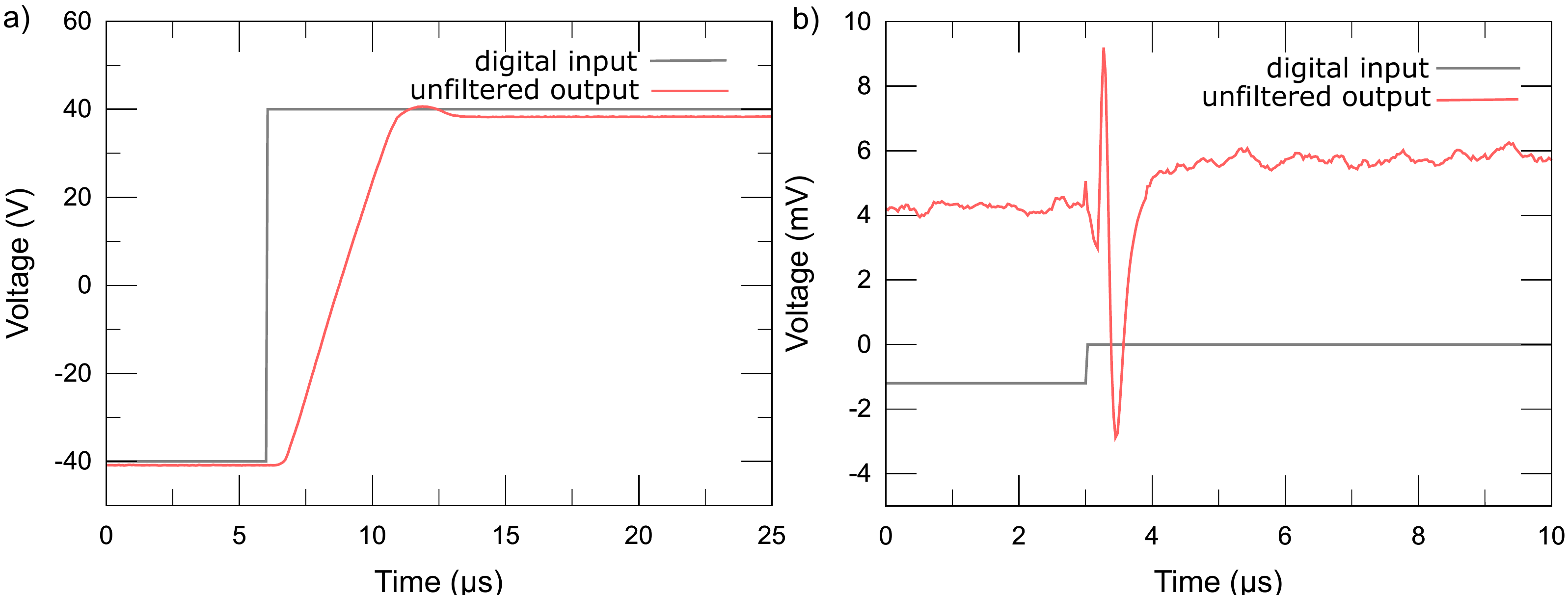}
	\caption{(a) Recorded full voltage swing from -40~V to 40~V. The initial and final settling behavior is due to trim capacitors that are employed to prevent overshooting and ringing at the output of the amplification stage. (b) Recording of a voltage glitch on the order of $\sim$10~mV occurring when the DAC switches all 16 output bits (1000 0000 0000 0000 0000 to 0111 1111 1111 1111).}
	\label{fig:bertha_StepGlitch}
\end{figure*}

An important parameter of the system is the \textit{slew rate}, which characterizes a limit on how fast the output voltage can be changed without causing a non-linear response. A full output voltage swing -40~V$\rightarrow$+40~V was recorded using a MSOX3104A mixed-signal oscilloscope and is shown in Fig. \ref{fig:bertha_StepGlitch} a). The maximum achieved slew rate is determined to be  $\sim\SI{20}{\volt\per\micro\second}$, consistent with the specified value for the LTC6090 operational amplifier used in the second amplification stage. As can be seen in Fig. \ref{fig:bertha_StepGlitch} a), the beginning and end of a full voltage swing display low-pass characteristics due to charging of additional trim capacitors. These trim capacitors (6.5 to 30~pF range) are placed between the negative input and output of the LTC6090 to prevent ringing at this amplification stage.  

\begin{table}
	\centering
\ifdefined\STIQPTwoCol
	\begin{tabular}{ | L{5.3cm} | C{1.6cm } | C{1.2cm} |}
\else
	\begin{tabular}{ | L{7cm} | C{2cm } | C{2cm} |}
\fi
    \hline
    \centering{\textbf{Characteristic}} & \textbf{Value} & \textbf{Unit} \\ \hhline{|=|=|=|}
    Noise level static output @ 1~MHz & $5\times 10^{-14}$ & V$^2$/Hz \\ \hline
    Max. glitch impulse area & 3(1) & nV$\cdot$s \\ \hline
  	Max. slew rate & 20 & V$/\mu$s  \\ \hline   
  	rms voltage stability & $\pm$ 140 & $\mu$V \\ \hline          
    \hline
    \end{tabular}
		\caption{Electrical performance characteristics of the mAWG.}
    \label{tab:bertha_noise}
\end{table}

Glitches, i.e. undesired voltage spikes, occur when the DAC switches to a different output voltage. Due to their short duration, they cover a wide frequency range and can lead to spurious motional excitation of the trapped ions upon propagation to the trap electrodes. Glitches are commonly characterized in terms of glitch impulse area, which is proportional to the number of bits flipped throughout the update. The measured glitch signal shown in Fig. \ref{fig:bertha_StepGlitch} (b) occurs during a programmed voltage change of one LSB at the mid-scale voltage output (0~V), which causes \textit{all bits} of the DAC to flip. This produces the worst-case glitch with an impulse area of 3(1)~nV$\cdot$s. Even if the static noise level of the mAWG would be sufficiently low to yield satisfactory heating rates, the presence of glitches makes additional low-pass filters a necessity. The glitch impulse area has also been measured using a MSOX3104A mixed-signal oscilloscope with a measurement bandwidth of 1~GHz and a resolution of 12~bits upon averaging.

Based on the measurements discussed above, a performance summary of the analog output modules is given in Tab. \ref{tab:bertha_noise}. In the following chapter \ref{sec:wiring}, the performance specifications will be analyzed in more detail, when the system is characterized in combination with an ion trap setup, including external filter stages and wiring.

\section{Routing of RF and DC signals to Ion Trap Apparatus} \label{sec:wiring}

Trapping ions in a segmented Paul trap requires large rf voltage amplitudes to provide confinement in the transverse directions and lower dc voltages to provide electrostatic confinement and control in the axial direction. In the following section we will discuss how large rf voltages, (e.g. 200~V amplitude at $\sim$35~MHz) are generated and applied to trap electrodes and how dc voltages generated by the previously discussed mAWG, section \ref{sec:bertha}, are filtered and connected to dc electrodes inside a UHV chamber. Furthermore we will discuss effects of inadequate DC and RF grounding and an additional noise analysis will be presented for a realistic experimental setup including all cables and filters.

\subsection{Trap Drive Routing and Stabilization}

Generating large rf voltages for the operation of a Paul trap commonly requires the use of a resonant circuit. Different resonator types depending on the ion trap type and ion species are employed. Low atomic mass ions are commonly trapped at higher trap drive frequencies on the order of up to 150~MHz, making a quarter-wave coaxial resonator a favorable choice \cite{Jefferts1995}. Operating a cryogenic ion trap apparatus close to liquid helium temperatures requires the resonator to be placed inside the vacuum chamber near the trap. This has lead to the development of superconducting helical resonators \cite{Poitzsch1996} and discrete component resonators using copper coils or superconducting coils \cite{Gandolfi2012,Brandl2016}. The most widely used resonator is the helical resonator \cite{Siverns2012}, which can be used for a variety of frequencies and achieves a high quality factor at room temperature ensuring a large rf voltage gain and high signal-to-noise ratio. 

The resonator is typically supplied by an rf signal generator via a low noise rf amplifier, often including an active amplitude stabilization circuit. For the microfabricated ion trap presented in Sec. \ref{sec:microtrap}, a low noise signal generator (R \& S SMB100), which is further amplified by a high-gain (46~dB) high-power (5~W) rf amplifier (minicircuits ZHL-5W-1) is connected to a home-built helical resonator. 

Temperature drifts, and drifting stray capacitances will cause trap drive amplitude fluctuations and therefore instabilities of the confining electric potential. As a result, the transverse secular frequencies will be unstable, which negatively effects the fidelity of gate operations, especially if two-qubit gates are implemented using the transverse modes, see Sec. \ref{sec:architecture}. To mitigate these fluctuations, an active stabilization system was developed, which employs a capacitive divider at the resonator output sampling the current rf amplitude. The signal is fed, via a rectifier, to a proportional-integral (PI) servo, which adjusts the signal generator output in real time, similar to the system described in \cite{Johnson2016}.

\begin{figure}
	\centering
	\begin{adjustbox}{max width=\columnwidth}
		\includegraphics{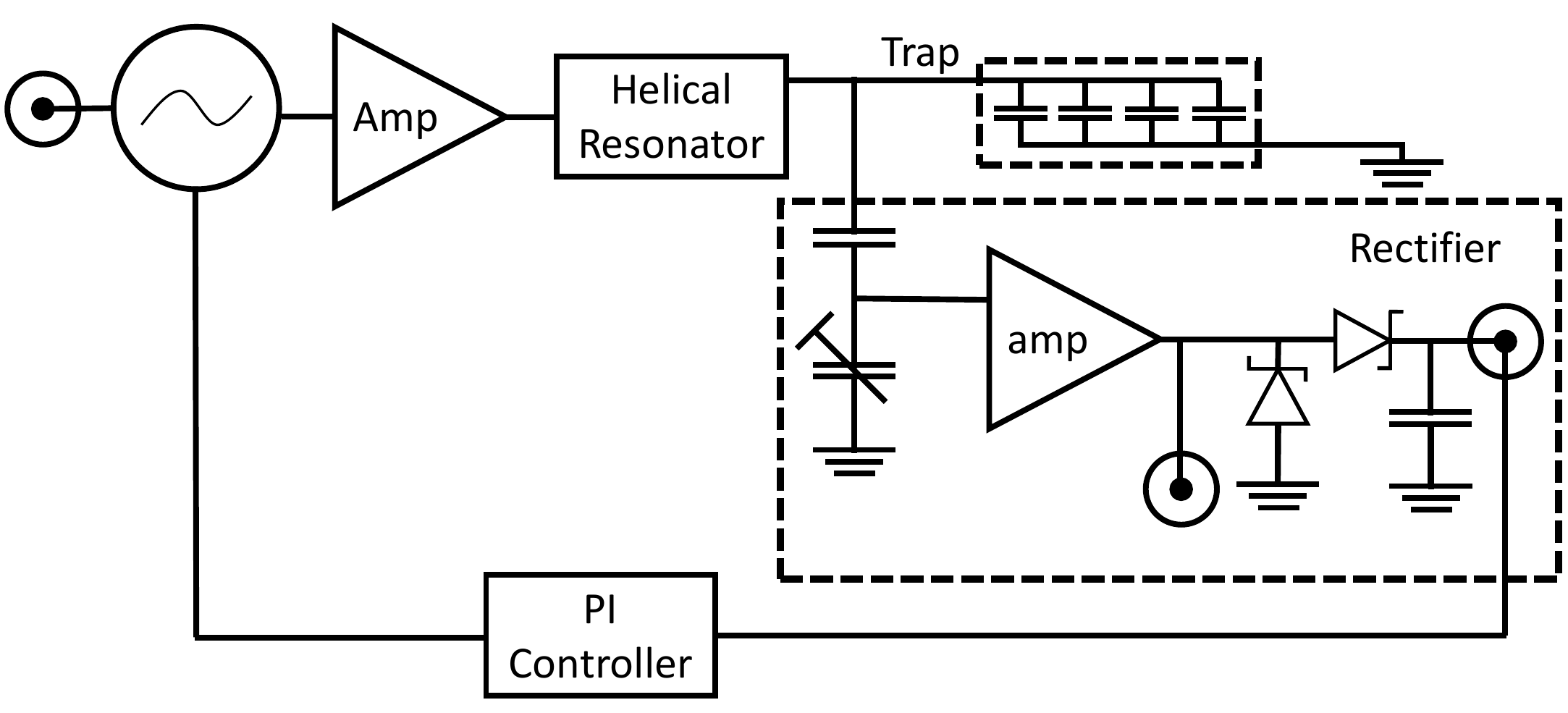}
	\end{adjustbox}
	\caption{Diagram showing the rf stabilization system, which uses a BNC connector to attach an adjustable capacitive divider to the resonant circuit. The operational amplifier buffers the picked up rf signal and decouples the stabilization circuit from the resonant circuit. The signal is then rectified and fed to a PI controller.} 
	\label{fig:wiring_rectifer}
\end{figure}

The challenge in active stabilization of the rf output amplitude lies in the high-impedance of the signal obtained from the capacitive divider, making the loop prone to electric noise pickup.  Our present design makes use of a capacitive divider, consisting of a 1~pF pickup capacitor and a 60-90~pF dividing capacitor, directly followed by an active rf buffer (TI BUF634). A schematic of this circuit is shown in Fig.\ref{fig:wiring_rectifer}. The buffer transforms the signal from the divider to a low impedance and prevents noise pickup within the stabilization loop. Fluctuations of the transverse secular frequencies due to the amplitude instabilities are on the order of $\sim 2\pi$~x~6~kHz without stabilization and as low as $\sim 2\pi$~x~20~Hz ($\sim 4\times 10^{-6}$ long-term stability) with active stabilization. Therefore, transverse mode secular frequency drifts do not represent a bottleneck for the fidelity of entangling gate operations. 

\subsection{Segment Voltage Filtering and Wiring}
\label{sec:filtersandwiring}

Technical noise, inherent to all electrical components, will have a detrimental effect on the system performance. Voltage noise at the secular frequencies leads to heating of the ions, and noise at lower frequencies will lead to motional decoherence, deteriorating the fidelity of entangling gate operations \cite{Turchette2000noise}. A lower bound for the heating rates is given by electric field noise generated by thermally activated processes at the trap surfaces \cite{Brownnutt2015, Boldin2018, Sedlacek2018, Deslauriers2006, Noel2019}.

Technical noise can be suppressed below the limit imposed by surface noise \cite{Sedlacek20182} by suitable filtering, however the shuttling-based quantum computing approach requires a sufficient bandwidth for fast operations, such that the heating rates are typically given by technical noise. 

A combination of low-pass filters is employed to reduce electric noise at frequencies below typical secular frequencies ($2\pi\times 1-7$~MHz) as much as possible. The cut-off frequency was chosen for sufficient suppression of noise at the secular frequencies, while still allowing for shuttling operations on timescales between 10 and 100~$\mu$s. Throughout separation and merging operations (see Sec. \ref{sec:separation}), the axial secular frequency transiently drops to a few 100~kHz, therefore a good rule of thumb is to choose the filter cut-off to be below the minimum secular frequency obtained throughout these operations. In our case, the minimal secular frequency during ion separation was determined to be 275~kHz and the cut-off frequency of the main RLC low pass filters was thus chosen to be 100~kHz.

\begin{figure}
	\centering
	\begin{adjustbox}{max width=0.45\textwidth}
		\includegraphics{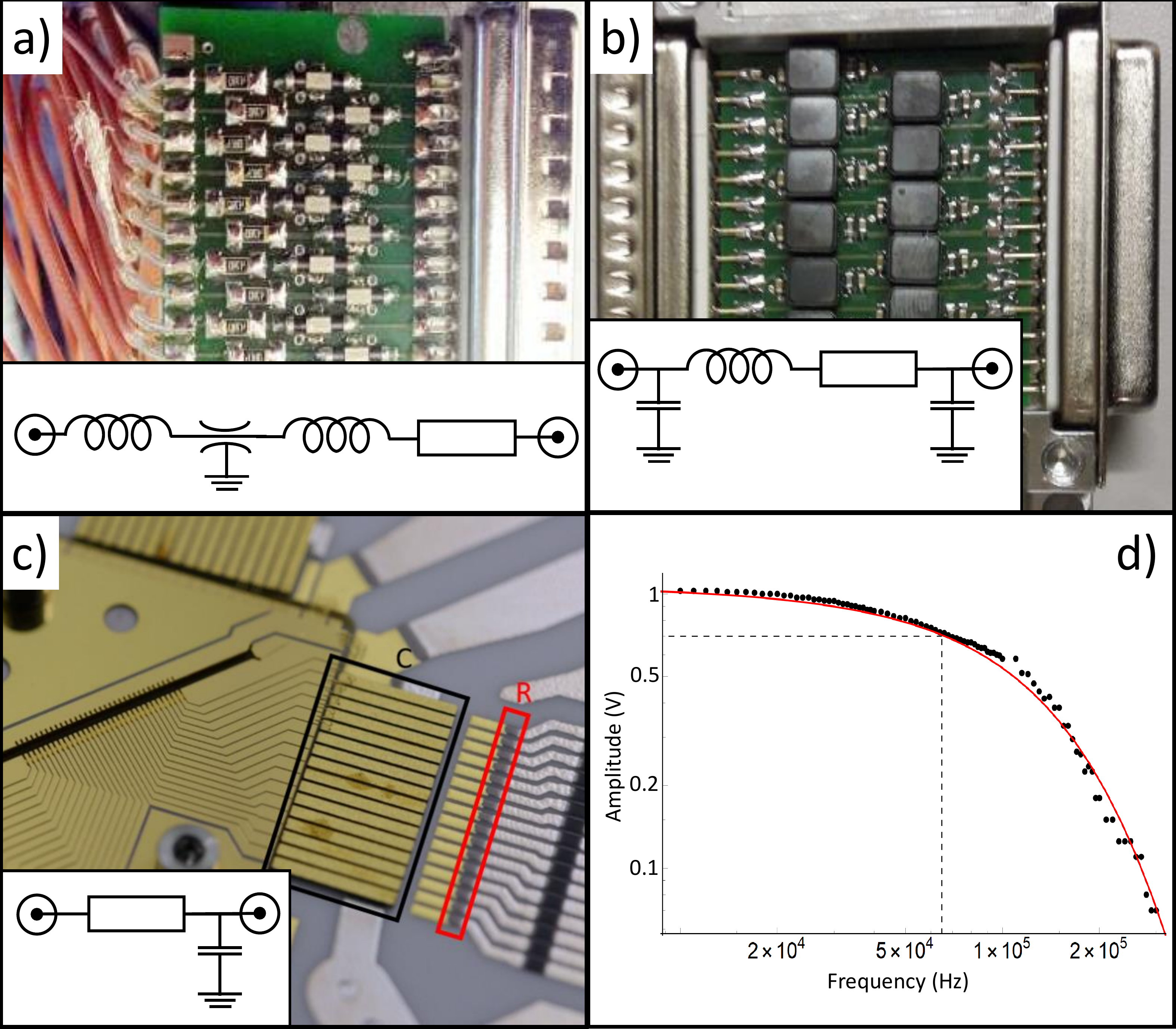}
	\end{adjustbox}
	\caption{Pictures showing the three filters placed between the mAWG and the ion trap electrode. The high frequency LC filter is shown in \textbf{a)}, made up of two ferrite core inductors and a 4.7~nF capacitor, followed by the main $\Pi$ filter shown in \textbf{b)}. These filters are made up of two 22~nF capacitors, a 150~$\mu$H inductor and a 69.8~$\Omega$ resistor. The last, in-vacuum filter stage is shown in \textbf{c)}. It is placed next to the trap and is comprised of a 10~$\Omega$ printed resistor and a 2~nF capacitor. \textbf{d)} Shows the filter response curve of the combined filters and wiring system. Dashed lines indicate the 3db point of the filter response curve.}
	\label{fig:wiring_LRCfilter}
\end{figure}

Three filter stages are employed in our system. First, a set of high frequency LC filters are placed at the outputs of the mAWG and used to prevent bleed-though of digital clock signals and other higher frequency noise from reaching the ion trap electrodes. These filters make use of a commercial LC electromagnetic interference (EMI) filter (NFE61PT472C1H9L) with a cut-off frequency of 400~kHz using a ferrite core based inductance and a capacitor. Ferrite cores are lossy at high frequencies and lead to dissipation of high-frequency signals. As shown in the schematic in Fig.\ref{fig:wiring_LRCfilter} (a), the filter can be effectively represented by two inductors, both sharing a common ferrite core, and a central capacitor of 4.7~nF. A total of 16 LC filters, shown in Fig.\ref{fig:wiring_LRCfilter} (a) with additional resistor are placed in a custom made D-Sub connector, which provides shielding and allows for small diameter RG178 coaxial cables to be directly soldered to the outputs of the high frequency filters. 

The second filter stage, possessing the lowest cut-off frequency, consists of PCB mounted PI-type filters in shielded metal boxes using D-sub connectors at the in and output. This makes it possible to easily exchange filters for a different cut-off frequency. The filters are directly plugged into the feedthroughs of the vacuum apparatus, see Fig.\ref{fig:wiring_LRCfilter} (b).  The filters consist of two 22~nF capacitors, a 150~$\mu$H inductor and a 69.8~$\Omega$ resistor per channel. The 3rd order filter is designed for a cut-off of 100~kHz. The inductors are shielded to reduce inductive coupling and thus cross-talk between neighboring channels. 

As a third and final filter stage, in-vacuum RC filters near the trap are used to reduce potential noise pick up inside the vacuum chamber and to help provide a low impedance path to ground for dc electrodes at the rf drive frequencies. 

The capacitances of these filters ensure a low impedance path to ground for the dc electrodes of the ion trap, which are capacitively coupled to the rf electrodes. 
If the filter capacitance would not be present, a significant rf voltage would drop at the resistive impedance of the dc supply lines. This would yield an increased effective load of the rf resonator, deteriorating its performance by decreasing the voltage gain and increasing its linewidth. Due to the reduced signal-to-noise ratio  heating rates due to noise on the rf signal, at frequencies offset from the trap drive frequency by the secular frequencies would increase. Furthermore, without a low impedance path to ground for the rf, the capacitively coupled dc electrodes would show a significant rf potential, instead of rf ground, compromising the confinement potential of the trap. 

In-vacuum RC filters are placed on the trap mount made from alumina and are shown in Fig.\ref{fig:wiring_LRCfilter} (c). The ultra-high vacuum environment requires careful choice of the filter components. The trap mount is made of alumina, and tracks are printed on it using sintered metal paste. The 10~$\Omega$ filter resistors are directly printed onto the alumina. Ultra-high vacuum compatible single layer ceramic capacitors arrays (2~nF) are soldered onto the metal tracks of the PCB for grounding and connected directly to the trap and signal tracks using wire bonding. These filters feature a cut-off frequency of 8~MHz, which only removes high frequency noise close to the drive frequency and above.   

Coaxial cables (RG178), are soldered to each output pin of the high frequency filters at the mAWG output and connected to the input pins of the PI-filters. They are bundled together and placed inside an additional metal mesh shielding shroud to further reduce noise pickup. Inside the vacuum chamber, space restrictions require the use of simple single-stranded wires without shielding to make a connection between the feedthrough and the RC filter board.

Careful design of the grounding topology can have a significant impact on the system performance in terms of heating rates and motional dephasing. Possible ground loops, i.e. spurious electrical potentials between circuit points intended to be ground, result in additional noise due to current flow between these points. The mAWG is equipped with an isolating transformer separating the mains ground from the ground of the power supplies. Analog grounds of the mAWG output modules are carefully routed to a common ground of the vacuum apparatus. The analog ground pins of the output modules are connected to the shielding of the coaxial cables  between output filters and PI-filters, which have a common ground for all channels. An additional shielding metal mesh is wrapped around the coaxial cables making contact between analog output ground and filter ground. Filter ground is directly connected to the vacuum chamber via the D-Sub connectors of the vacuum feedthroughs.  Vacuum apparatus ground is finally connected to the laboratory ground using low resistance connections. 

Filter responses and cut-off frequencies of individual filters are only of limited use since a filter or even the cables can influence the behavior of the consecutive filter and therefore the entire system. We have thus measured a filter response curve of the combined filters and wiring system placed between the mAWG and ion trap. The results of this measurement are show in Fig. \ref{fig:wiring_LRCfilter} d). The combined filter and wiring setup has a significantly lower cut-off as compared to the dominating PI-filter. This is due the 50~$\Omega$ resistor after the first LC filter, which is in series with the 69.8~$\Omega$ resistor of the PI-filter.

\subsection{Filter Response and Technical Performance Analysis of the Setup}

In this section, we discuss the effect of the main PI-filter stage on the static noise on the dc lines and the glitches occurring upon DAC updates.

\begin{figure}
	\centering
	\includegraphics[width=0.45\textwidth]{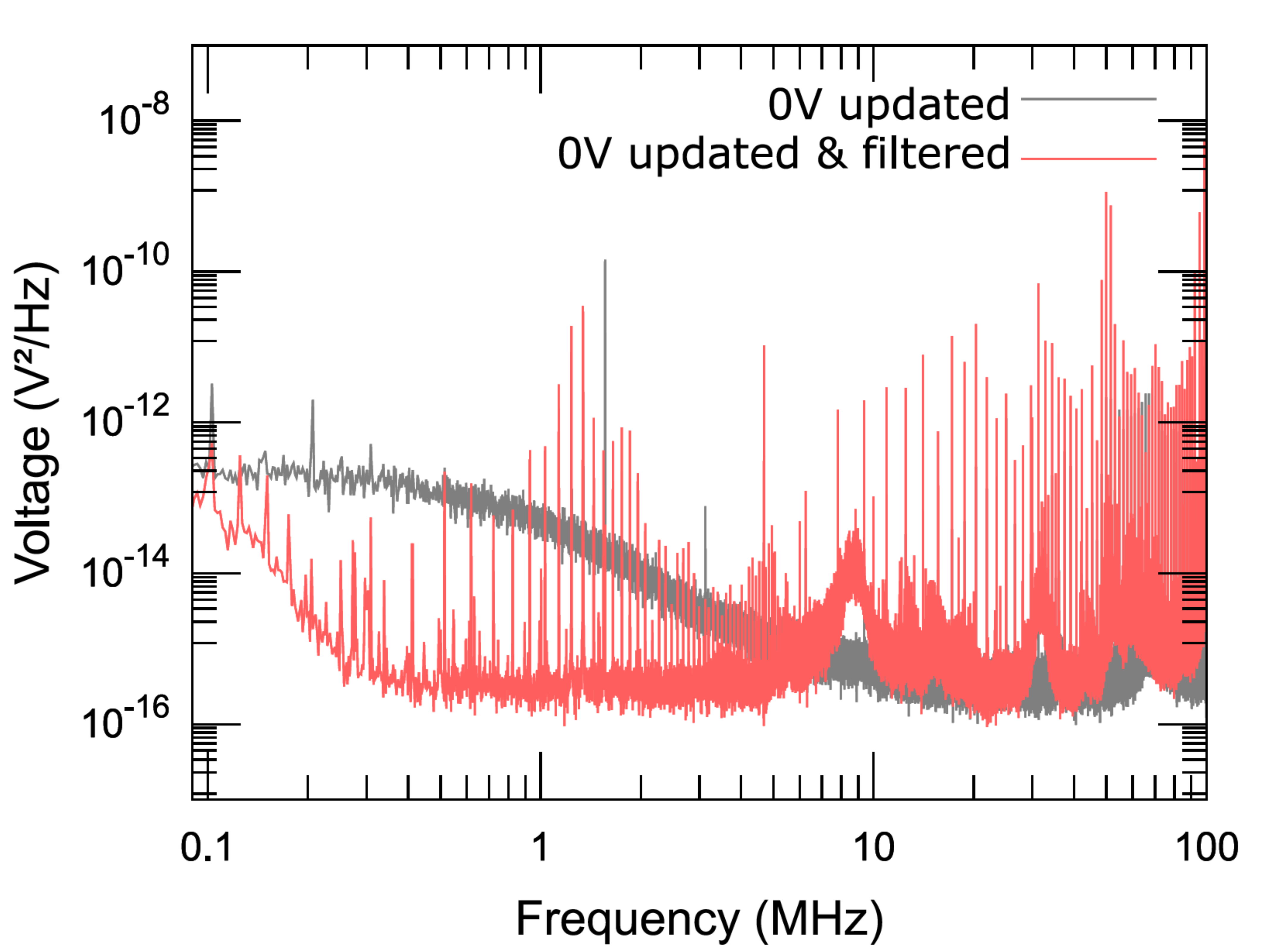}
	\caption{Measured noise spectral densities for 0~V output at continuous DAC updates. The gray curve is also shown in Fig. \ref{fig:bertha_noise} and corresponds to measurement at the mAWG output, and the red curve corresponds to measurement at the PI-filter output. As the measurement is performed on the bare filter output pins without proper shielding, noise pickup occurs and manifests in sharp peaks. The filtered signal reaches the measurement noise floor between 300~kHz and 5~MHz, compare to Fig. \ref{fig:bertha_noise}. The actual noise suppression by the PI-filters is presumably better.}
	\label{fig:wiring_filterNoise}
\end{figure}

Measurements of the mAWG output noise shown in Sec. \ref{sec:bertha_mAWG Noise Spektrum} indicate that additional filtering is required to achieve sufficiently low heating rates.  Noise spectral densities measured at the output of the PI-filter stage are shown in Fig. \ref{fig:wiring_filterNoise}. The measurement has been performed identically to the noise measurements shown in Sec. \ref{sec:bertha_mAWG Noise Spektrum}. The output channel has been set to 0~V, and output noise with continuous DAC updates is shown in comparison to the measurement directly at the mAWG output. In the range between 100~kHz and about 5~MHz, noise suppression by the filters is clearly visible. However, the measurement is limited by the noise floor of the used spectrum analyzer of about $\SI{10e-16}{\volt^2\per\hertz}$, this measurement scheme does therefore not allow for a quantitative prediction of the technical noise contribution to ion heating rates, see Sec. \ref{sec:heating}. Spurious peaks can be seen in the spectrum, these presumable occur because the measurement is prone to electromagnetic pickup, such that the noise peaks do not necessarily occur in the trap setup. The broad peak slightly below $\SI{10}{\mega\hertz}$ corresponds to the self resonance of the inductor used for the PI-filters.

\begin{figure}
	\centering
	\includegraphics[width=0.4\textwidth]{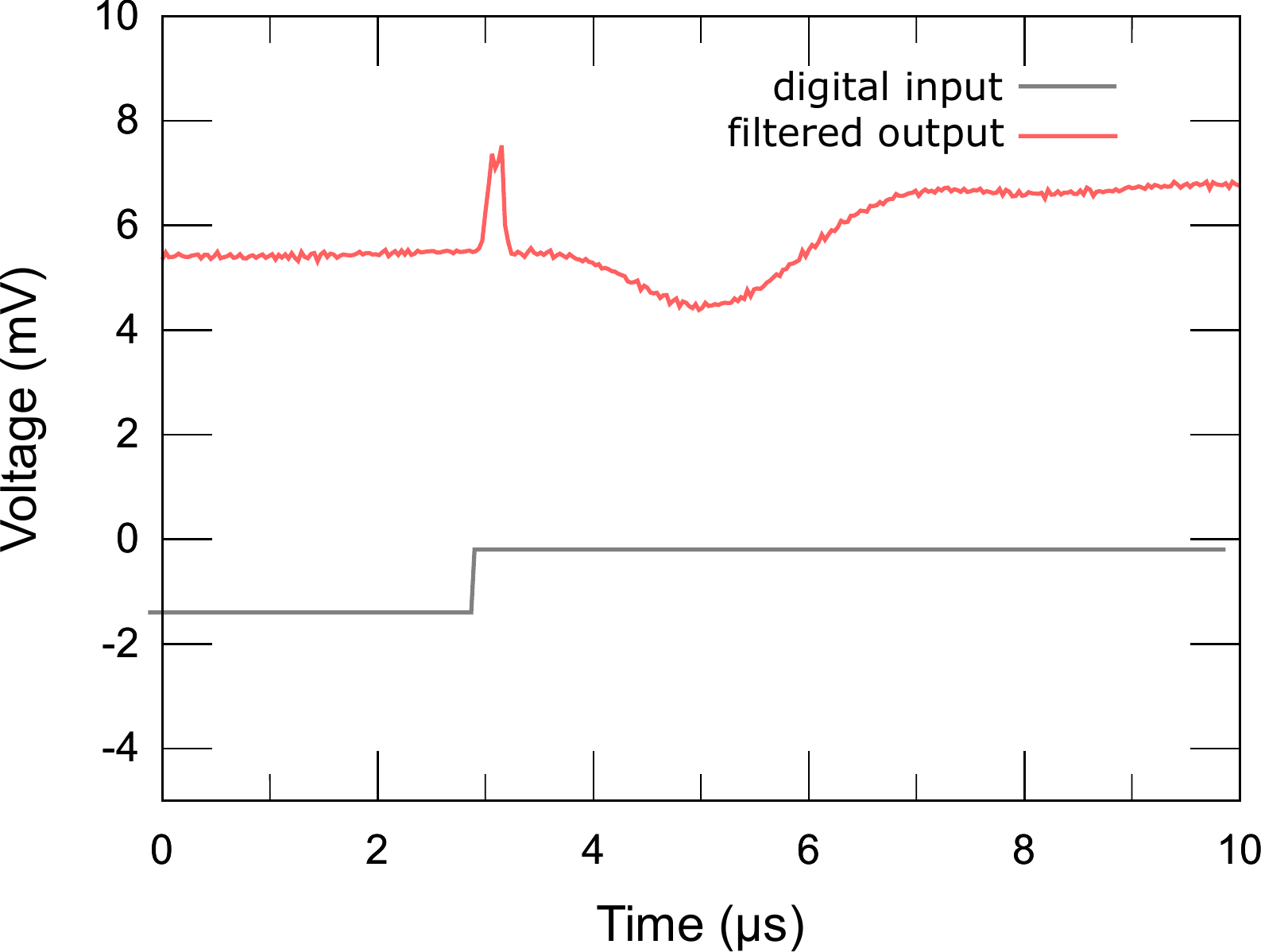}
	\caption{ Voltage glitch occurring when the DAC switches all 16 output bits (1000 0000 0000 0000 0000 to 0111 1111 1111 1111), affected by the filter system. The glitch amplitude is reduced by a factor of $\sim$4 and significant smearing compared to unfiltered glitches can be observed.}
	\label{fig:wiring_StepGlitch}
\end{figure}

The voltage glitches observed when switching DAC bits are of broadband nature, see Sec. \ref{sec:bertha_mAWG Noise Spektrum}, and will therefore be mitigated by the low pass filters. Significant suppression of the glitches can be seen in Fig.\ref{fig:wiring_StepGlitch} in comparison to Fig. \ref{fig:bertha_StepGlitch}. Glitch energy can not be entirely removed from the system since the low pass filters charge up during filtering and slowly dissipate the energy after the glitch has passed, a phenomenon commonly called glitch smearing.

\section{System characterization}\label{sec:characterization}

The electric field noise affecting the trapped-ion qubits is a key limitation to the performance of the architecture. It stems from molecular processes on the trap surfaces \cite{Talukdar2016, Noel2019} and from the electrical circuitry supplying the dc and rf trapping voltages. While the latter -technical- noise can in principle be filtered, the shuttling-based trapped-ion QC approach requires supplying a bandwidth sufficient for fast shuttling operations to the dc electrodes, such that technical noise will typically limit the system performance. Electric field noise manifests in two detrimental effects: First, noise at the secular frequencies causes heating of the stored ions \cite{Brownnutt2015}, resulting in thermal population of the secular modes, which leads to decreased entangling gate fidelities. Second, noise ranging from slowly varying secular frequency offsets to secular frequency modulation at up to a few 100~kHz leads to decoherence of transient motional superposition states during gate operations, which also represents an error source.\\
Therefore, our architecture requires sufficiently low noise on the trap electrodes across the entire range up to the secular frequencies, and in addition a suitable tradeoff has to be made between the bandwidth of the dc signals allowing for fast shuttling operations and the overall noise level suppression. Here, we present the resulting performance figures in terms of heating rates and motional coherence.

\noindent
\subsubsection{Ion heating}
\label{sec:heating}

\begin{figure}
	\centering
	\includegraphics[width=0.45\textwidth]{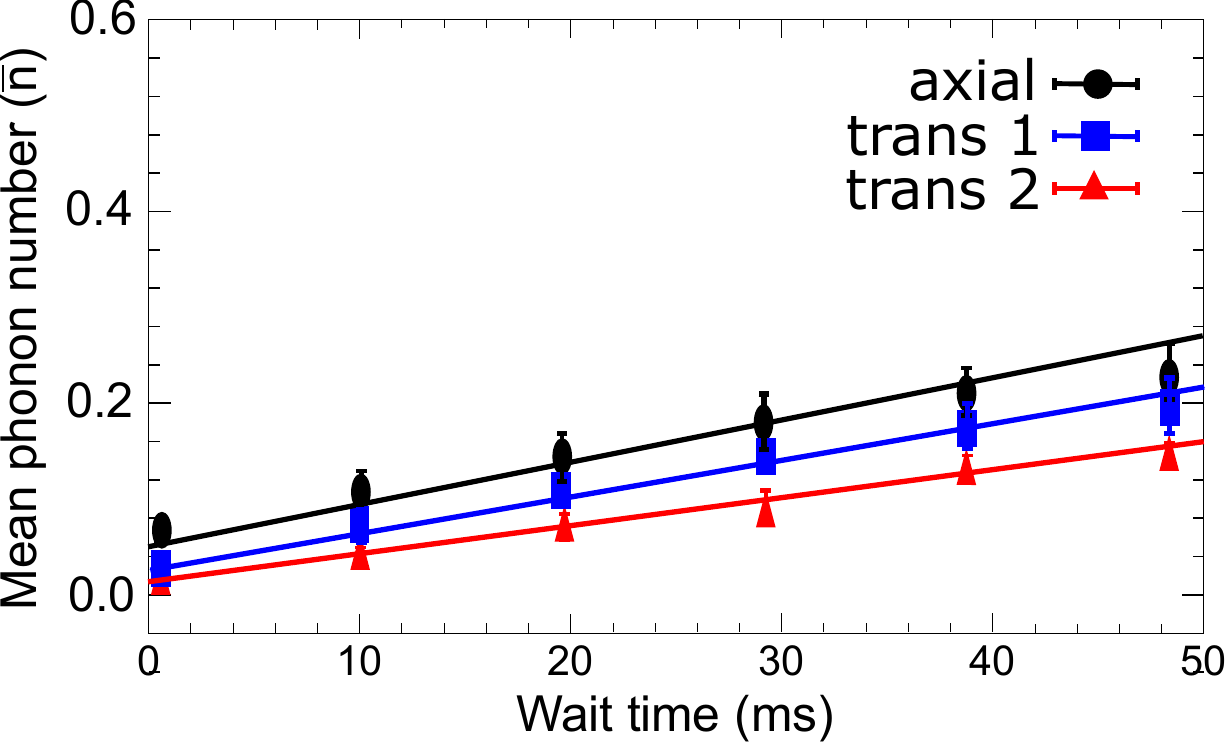}
	\caption{Heating rates: Measured mean thermal phonon numbers on all secular modes of a single ion with variable wait times after sideband cooling are shown with an  external low-pass filters at a cut-off frequency $\SI{100}{\kilo\hertz}$. Heating rates are 5.5(3) phonons per second for the axial mode, 4.8(2) for the transverse mode 1 and 3.6(1) for transverse mode 2.}
	\label{fig:heatingrate}
\end{figure}

The heating rates on all three secular modes are measured for a single ion in the laser interaction zone (LIZ), i.e. the trap segment to which all laser beams are directed. The measurements utilize the Rabi flopping method as described in Sec. \ref{sec:thermometry}.
Here, a variable wait time is introduced between sideband cooling close to the ground state of all modes and the probe pulse. The inferred mean thermal phonon numbers versus wait time are shown in Fig. \ref{fig:heatingrate}. On the dc lines, PI-type low-pass filters at a cut-off frequency of $\SI{100}{\kilo\hertz}$ were employed, see Sec. \ref{sec:wiring}. The resulting heating rates are 5.5(3) phonons per second for the axial mode at $2\pi\times$~1.49~MHz, 4.8(2) for the transverse mode 1 at $2\pi\times$~3.88~MHz and 3.6(1) for transverse mode 2 at $2\pi\times$~4.66~MHz. More sets of heating data have been acquired for different trap voltages, i.e. different secular frequencies. It was found that the dependence of the heating rates on the secular frequencies is not consistent with a proper power-law behavior, as would be expected for anomalous heating originating from surface heating \cite{An2019}. This indicates that at least the transverse modes are predominantly affected by technical noise. This is supported by the measured noise spectrum at the output of the $\Pi$ filter, Fig. \ref{fig:wiring_filterNoise}, which exhibits many resonances in the range of typical secular frequencies.

This unusual behavior can be explained by the significant voltage noise peaks observed when the PI-type low-pass filters is connected to the mAWG show in Fig. \ref{fig:wiring_filterNoise} and discussed in section \ref{sec:wiring}. These noise peaks can overlap with the secular frequency or a combination of the secular frequency and drive frequency. This will lead to a significant increase in technical noise, which can dominate the anomalous heating for that mode.

\subsubsection{Motional dephasing} 
\label{sec:motionaldephasing}
\begin{figure}
	\centering
	\includegraphics[width=0.45\textwidth]{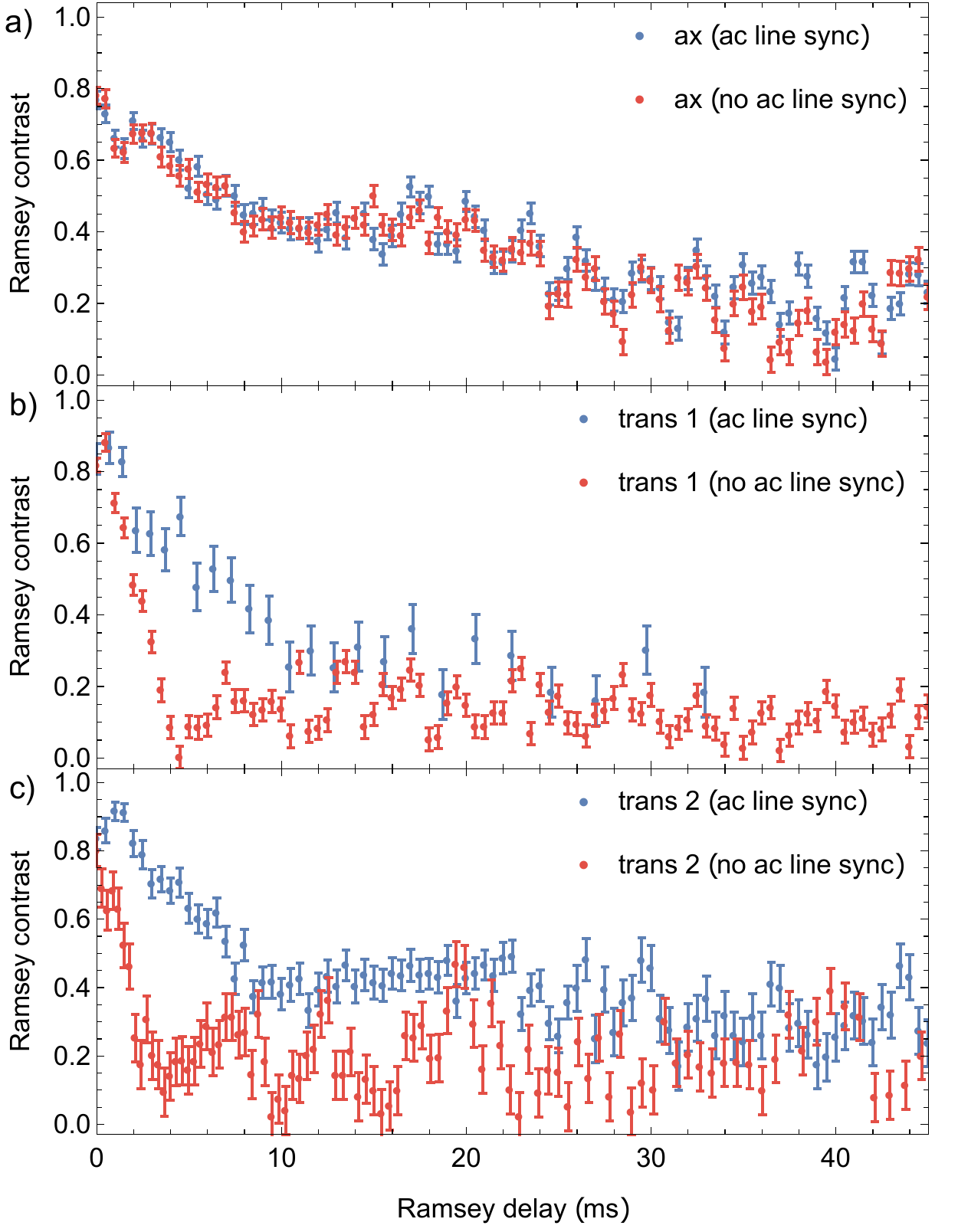}
	\caption{Motional coherence decay. We measure the decay of  superposition state coherence for all three secular modes of a single ion (see text). Data is shown for the cases with (without) triggering the measurement to the ac mains in blue (red), for the axial secular mode in \textbf{a)}, the lower-frequency transverse mode in \textbf{b)} and the higher-frequency transverse mode in \textbf{c)}.}
	\label{fig:motionalcoherence}
\end{figure}

Noisy dc and rf trap voltages lead to fluctuating secular trap frequencies, which in turn lead to dephasing of motional superposition states occurring during entangling gate operations \cite{Turchette2000noise, Ballance2014}. White secular frequency noise leads to exponential decay of motional coherences at a rate $\gamma_m$. We measure these rates on the transverse secular modes of a single ion in the motional ground state in the LIZ by means of a motional Ramsey scheme. We proceed in the following steps: i) A qubit superposition $\left(\ket{\uparrow}+\ket{\downarrow}\right)\ket{0}$ is created by a resonant $\pi/2$ qubit rotation, ii) a $\pi$ rotation on the red sideband of the mode of interest creates the superposition $\ket{\uparrow}\left(\ket{0}+\ket{1}\right)$, iii) dephasing takes place during a variable wait time $T$, iv) another $\pi$ rotation maps the motional superposition back into a spin superposition, and v) a concluding $\pi/2$ spin rotation with(without) $\pi/2$ phase shift serve for measurement of the spin $\hat{\sigma}_X$($\hat{\sigma}_Y$) operators. The resulting measured Ramsey contrast is then obtained from $\mathcal{C}(T)=\sqrt{\langle\hat{\sigma}_X\rangle^2+\langle\hat{\sigma}_Y\rangle^2}$ and decreases with $T$. The results are shown in Fig. \ref{fig:motionalcoherence}.

In order to verify the presence of 50~Hz ac line pickup on the trap electrodes, the measurements were carried out with and without triggering to the ac mains signal. While no difference is observed for the axial mode, ac line triggering yields substantially longer coherence times on both transverse modes. This result indicates either ac mains pickup in the rf system or ac mains common-mode noise on the dc lines. Entangling gates will suffer errors from motional dephasing of magnitude $\lesssim t_g \gamma_m$, where $t_g$ is the gate duration \cite{Ballance2014}. We estimate the error contribution from this effect to be smaller than 0.25\% for gate of $t_g=\SI{50}{\micro\second}$ driven on the higher-frequency transverse mode. This shows that this gate error source -which is often disregarded- can contribute significantly to gate error budgets. We emphasize the importance of measuring motional dephasing rates along with heating rates when the performance of a trap setup is to be characterized.

\section{Shuttling}\label{sec:shuttling}

Shuttling operations are executed by applying varying voltage potentials to the dc electrodes, leading to a change of trapping potentials along the axial direction over time. For example, we consider a single ion confined in a harmonic electrostatic potential well generated by a negative trapping voltage applied to segment $n$ while all other segments are kept at 0~V. Ramping the voltage at segment $n$ to 0~V while simultaneously ramping the neighboring segment $n+1$ to a negative voltage, the electrostatic potential well will shift along the trap axis, from segment $n$ to $n+1$. The ion confined within this well will move along with respect to the center of the potential well. This corresponds to a pendulum with dragged support, which can lead to oscillatory excitation along the direction of transport, increasing at short process times.\\

As explained in Sec. \ref{sec:architecture}, shuttling operations should take place as fast as possible, while motional excitation is to be avoided. In our architecture, entangling gates and readout are only sensitive to transverse motion, see Sec. \ref{sec:architecture}. Therefore, the predominantly axial excitation can be tolerated to some extent. However, excessive axial excitation amplitudes are to be avoided, as there is i) residual coupling to axial motion throughout gates and readout, ii) anharmonic coupling between axial and radial collective modes, iii) large axial excitation can render separation/merging operations unreliable and iv) oscillation amplitudes comparable to laser focus size lead to undesired intensity modulation during gate operations. Importantly, as the trap geometry is non-ideal, i.e. not entirely uniform and symmetric. Therefore, shuttling along the the trap axis also leads to parasitic transverse excitation. 

In order to realize the desired operational principle of the quantum processing node, sequences for reconfiguration of the qubit register have to be carried out, which consist of three elementary shuttling operations, see Fig. \ref{fig:architecture_overview}:
\begin{itemize}
\item
\textbf{Linear transport:} Movement of single ions or small ion crystals along the trap axis.
\item
\textbf{Separation/merging:} Ion crystals are split into single ions and/or smaller crystals, or the reverse process is carried out. 
\item
\textbf{Swapping:} Physical swapping through rotation of ion crystals, equivalent to a logical SWAP gate.
\end{itemize}

In the following, we describe the measurement methods employed for characterizing the performance of the shuttling operations in Sec. \ref{sec:thermometry}, then we cover the operational building blocks linear transport (Sec. \ref{sec:transport}), separation / merging (Sec. \ref{sec:separation}) and ion crystal rotation (Sec. \ref{sec:swapping}).

\subsection{Excitation measurement}\label{sec:thermometry}

In order to verify and optimize shuttling operations, suitable measurement schemes for characterizing the state of a given motional mode of a single ion or ion crystal is required. A well-established method consists of recording Rabi flops on a coherently driven transition between long-lived internal states of the ions, which feature a suitable coupling to the motional mode(s) of interest \cite{Meekhof1996, Leibfied2003rmp}. In our case, we utilize stimulated Raman transitions \cite{Wineland1998} between the Zeeman sublevels of the S${1/2}$ ground state of $^{40}$Ca$^+$, i.e. $\ket{m_J=-1/2}\equiv\ket{\downarrow}$ and $\ket{m_J=+1/2}\equiv\ket{\uparrow}$. These are driven by laser beam pairs, which are directed such that the effective wavevectors point either along the axial direction or along the transverse directions. The resulting Lamb-Dicke factors, characterizing the coupling to the motion, range from about 0.05 for the lower-frequency transverse center-of-mass mode of a two-ion crystal, up to about 0.24 for the axial vibration of a single ion. Note that such measurements can also be performed by coherently driving electronic transitions which are dipole-forbidden, such as the S$_{1/2}\leftrightarrow$D$_{5/2}$ electric quadrupole transition in alkaline-earth ions \cite{Roos1999}.

The shuttling operations predominantly cause coherent oscillatory motion, therefore one is not interested in full quantum state reconstruction, but rather in obtaining the mean phonon number $\bar{n}=\vert\alpha\vert^2$ pertaining to the displacement parameter $\alpha$. Resonantly driving Rabi flops on a sideband transition where a $\pi$ rotation (a spin flip in our case) is accompanied by a change of $\delta n$ phonons, starting with a single ion in $\ket{\downarrow}$, the probability to find the ion in $\ket{\uparrow}$ after exposure time $t$ is given by

\begin{equation}
P^{(\delta n)}_{\ket{\uparrow}}(t)=\sum_{n=0}^{n_{\text{max}}} p_n(\alpha) \frac{1}{2}\left(1-\cos\left(\Omega_{n,n+\delta n} t\right)\right)
\label{eq:rabiflops}
\end{equation}

Here, $p_n(\alpha)=e^{-\vert\alpha\vert^2}\frac{\vert\alpha\vert^{2n}}{n!}$ is the Poissonian phonon number distribution pertaining to oscillatory excitation, and Rabi frequencies $\Omega_{n,n+\delta n}$ depend on the phonon number via \cite{Wineland1998}

\begin{eqnarray}
\vert\Omega_{n,n+\delta n}\vert&=&\Omega_0 \eta^{\delta n} e^{-\eta^2/2}\sqrt{\frac{n_<!}{n_>!}} L_{n<}^{\delta n}(\eta^2) \nonumber \\ &\approx& \begin{cases} \eta\sqrt{n} & \delta n=-1 \\ 1-\eta^2 n & \delta n=0 \\ \eta\sqrt{n+1} & \delta n=+1 \end{cases},
\end{eqnarray}

where $\Omega_0$ is the base Rabi frequency, $\eta$ is the Lamb-Dicke factor, $n_< (n_>)$ is the smaller (larger) number of $\{n,n+\delta n\}$ and $L_n^m(x)$ are the generalized Laguerre polynomials. The approximation in the second line holds within the Lamb-Dicke regime $\eta n \ll 1$, where the spatial extent of the wavefunction is much smaller than the wavelength of the driving radiation. The three most common cases are the first red sideband $(\delta n=-1)$, the first blue sideband $(\delta n=+1)$ and the carrier transition $(\delta n=0)$. For the characterization of heating rates, as shown in Fig. \ref{fig:heatingrate}, the Poissonian phonon number distribution in Eq. \ref{eq:rabiflops} has to be replaced with a thermal distribution $p_n(\bar{n})=\bar{n}^n/(\bar{n}+1)^{n+1}$, with the mean thermal phonon number $\bar{n}$.

In order to increase the accuracy of such measurements, Rabi flops can be recorded on several sidebands, which also serves for fixing parameters such as the base Rabi frequency $\Omega_0$ or the Lamb-Dicke factor. The model Eq. \ref{eq:rabiflops} can be straightforwardly extended to the presence of spectator modes \cite{Kaufmann2017swap}, which is required e.g. for two-ion crystals or for characterizing the state of transverse secular modes, where both modes for a single ion feature nonzero Lamb-Dicke parameters. Moreover, the model can be extended to include thermal excitation as well, which is required to characterize separation processes at long durations \cite{Ruster2014}. \\

Precise measurement of motional amplitudes using the method described above requires taking large amounts of data - typically $10^4-10^5$ individual measurements are required to infer motional excitation at a resolution down to 0.1 phonons. Therefore, for efficient calibration of the shuttling operations, a more convenient method is needed. In \cite{Walther2012}, a suitable metric coined \textit{pseudo-energy} was introduced, which has the following properties: i) It should scale monotonously with the mean phonon number up to an upper limit, ii) it should assume zero for $\bar{n}=0$ and iii) should exhibit an almost constant slope with respect to $\bar{n}$ across the region of validity. Such metric can be constructed as the weighted sum of a set of spin readout results pertaining to different sideband transitions $\delta n^{(i)}$ and probe times $t_i$. With the parameters being tailored for a specific purpose, such a measure can serve e.g. to efficiently minimize shuttling-induced excitation with respect to given control parameters.\\

\subsubsection{Simplified measurement scheme}\label{sec:thermometrysimple}
A rather simple and robust measurement scheme can be applied for small excitations, i.e. values of $\bar{n}$ for which $p_0(\alpha)\gg 0$ holds. Probing a red sideband for a range of time values $t$ and averaging the resulting values $P^{(\delta n)}_{\ket{\uparrow}}(t)$ is equivalent to averaging Eq. \ref{eq:rabiflops} over time. For all terms except $n=0$, the right-hand-side of Eq. \ref{eq:rabiflops} yields a constant:
\begin{eqnarray}
\overline{P^{(\delta n)}_{\ket{\uparrow}}}&=&\frac{1}{2}\sum_{n=1}^{n_{\text{max}}} p_n(\alpha)\approx \frac{1}{2}\left(1-p_0(\alpha)\right) \nonumber \\
\Rightarrow \bar{n}&=&-\ln\left(1-2 \overline{P^{(\delta n)}_{\ket{\uparrow}}}\right).
\label{eq:nbarsimple}
\end{eqnarray}
Therefore, sufficiently small values for $\bar{n}$ pertaining to a Poissonian phonon number distribution can be estimated from averaging recorded red sideband excitation probabilities over time. Eq. \ref{eq:nbarsimple} also holds for the case of coupling to multiple motional modes and in the presence of additional dephasing, and can therefore conveniently be applied for characterizing oscillatory excitation on multi-ion crystals.

\subsection{Linear transport}\label{sec:transport}

The first demonstration of transport in a segmented trap was reported in \cite{Rowe2002}. In \cite{Blakestad2009}, it has been shown that qubit states are preserved during transport, and in \cite{BLAKESTAD2011}, adiabatic transport retaining the ground state of axial motion has been demonstrated. In \cite{Walther2012,Bowler2012}, \textit{diabatic} transport has been shown: Due to the (coherent) oscillatory nature of the transport-induced excitation, it is possible to control the residual excitation either by choosing 'sweet-spot' transport times or by removing the excitation with a 'kick' voltage impulse. However, these approaches are of limited use for a scalable quantum processing node, as both require precise calibration of parameters such as a kick voltage amplitude, which might vary with the location of the transport and even drift with time.\\
Several control strategies have been proposed for robust control of transport induced excitation, such as optimal control \cite{Schulz2006} or invariant-based inverse engineering \cite{Torrontegui2011,Lu2014}. It has been shown in \cite{Fuerst2014} that in the limit of infinite resources such as segment voltage range and bandwidth, the control strategies yield a straight-forward bang-bang type solution.

If we move the center of a harmonic potential well $z_0$ pertaining to constant axial secular frequency $\omega$ along the trap ($z$) axis such that it follows the trajectory $z_0(t)$, the resulting displacement for a single ion of mass $m$ after the transport of duration $T$ can be calculated \cite{Bowler2012}:

\begin{equation}
\alpha(T)=-e^{-i\omega T}\sqrt{\frac{m\omega}{2\hbar}}\int_0^T \dot{z}_0(t') e^{i\omega t'} dt'.
\end{equation}

\begin{figure*}
	\centering
	\includegraphics[width=0.7\textwidth]{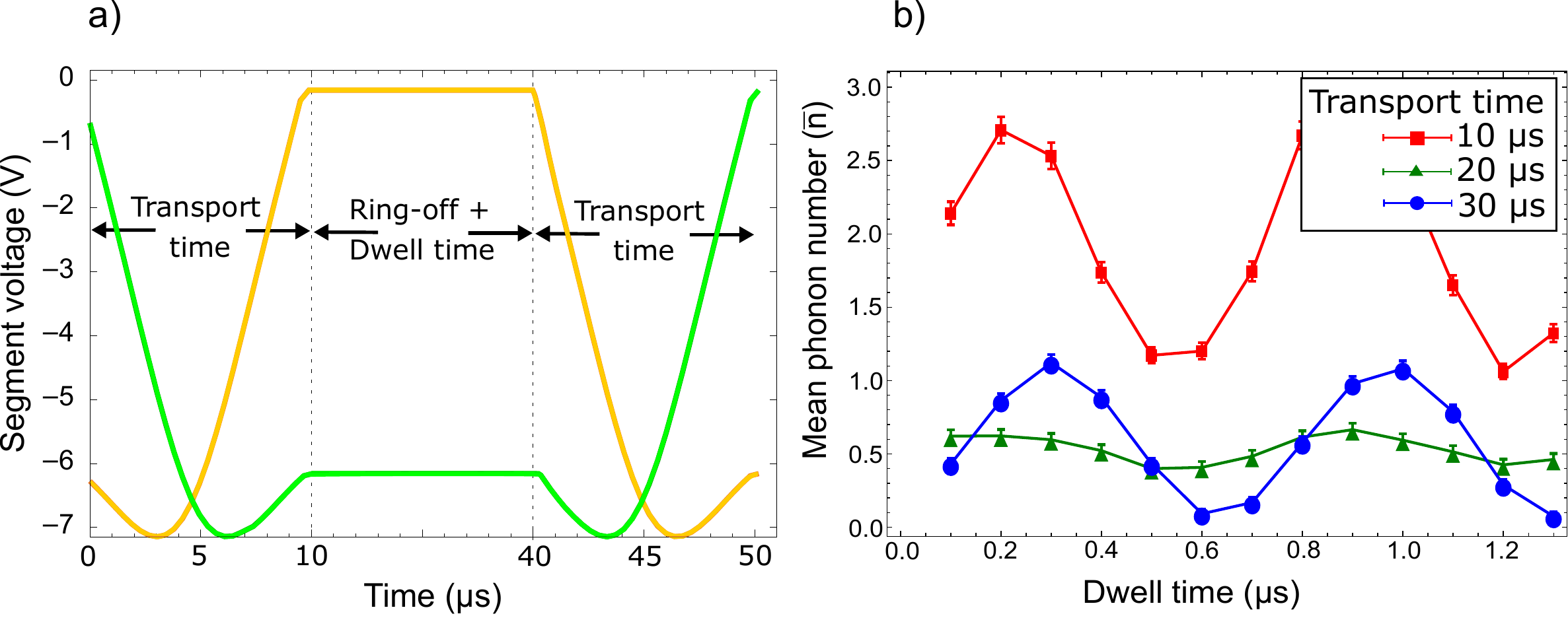}
	\caption{Linear transport. \textbf{a)} shows the voltage ramps for shuttling an ion from the LIZ to a neighboring segment and back. The yellow (green) curve shows voltage applied to the initial (destination) electrodes during the transport time. The ion is kept at the destination segment for a ring-off time of 30$\mu$s and a variable dwell time. \textbf{b)} shows the measured motional excitation on the axial mode resulting from this round-trip transport, versus the dwell time at the destination for multiple transport times. The coherent oscillatory nature of the excitation is revealed by the oscillations of the excitation with respect to the dwell time. It can be seen that excitations in the few-phonon range occur.}
	\label{fig:transportaxialexcitation}
\end{figure*}

This means that the magnitude of the residual excitation is given by the Fourier component of the trap center's velocity at the trap frequency. As our architecture allows for performing probing and readout only at the LIZ, we characterize the residual axial excitation by means of a round-trip scheme: A single ion is cooled close to the ground state of its axial motion. It is then shuttled from the LIZ to a neighboring segment, where it is kept idle for a variable dwell time, after which it is shuttled back to the LIZ. There, the state of its axial motion is probed via the Rabi flopping method described in Sec. \ref{sec:thermometry}. The employed voltage ramps are shown in Fig. \ref{fig:transportaxialexcitation} a), and the measured axial excitation for different shuttling durations is shown in Fig. \ref{fig:transportaxialexcitation} b). The voltage ramps are chosen such that a smooth acceleration and deceleration is ensured, minimizing the Fourier component of the well center's velocity $\dot{z}_0$, and to keep the axial secular frequency $\omega\approx$ $2\pi \times 1.5$MHz roughly constant. During the dwell time $t_D$, the oscillatory motion leads to accumulation of a phase $t_D \omega$, and the transport back to the LIZ leads to displacement of the same magnitude as the transport forth, but at reversed phase. The dwell time phase thus affects the magnitude of the final displacement, leading its oscillatory behavior with respect to $t_D$. The maximum excitation values therefore pertain to addition of the oscillatory excitation resulting from the back and forth transports, while the minimum values indicate a mismatch of these magnitudes: Because of filter-induced voltage waveform distortions, the voltage ramps for the two transport operations are not exactly symmetric, such that the excitation from the transport back cannot exactly null the excitation from the transport forth.

\begin{figure*}
	\centering
	\includegraphics[width=0.7\textwidth]{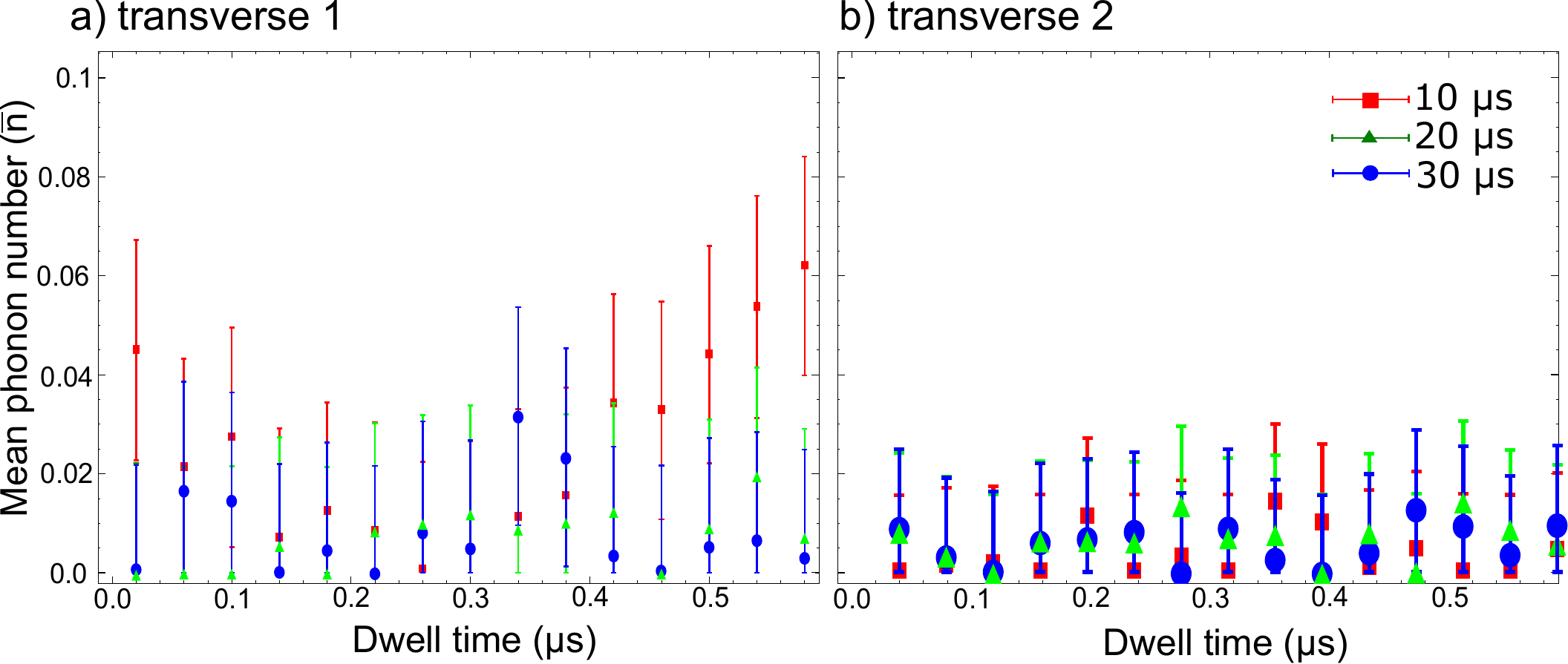}
	\caption{Transport-induced transverse excitation. \textbf{a)} shows the excitation on the lower-frequency transverse mode after a round trip transport, and \textbf{b)} shows similar data for the higher-frequency transverse mode. Data for various single-trip transport times is shown for varying dwell times. Small transverse excitation only occurs on the lower-frequency transverse mode.}
	\label{fig:transportradialexcitation}
\end{figure*}

The resulting values in the few-phonon range for transport durations of a few ten microseconds are tolerable, as all relevant laser-driven operations are only sensitive to transverse motion, see Sec. \ref{sec:architecture}. We probe the transverse excitation with the same scheme as described above, however we probe Rabi oscillation on sidebands pertaining to the lower- and higher-frequency transverse modes. The results are shown in Fig. \ref{fig:transportradialexcitation}. Residual transverse excitation can occur due to technical imperfections, i.e. if slightly different filtering characteristics on the dc lines pertaining to two adjacent dc segments lead to spurious transient transverse electric fields, if trap geometry imperfections lead to slightly tilted confining potentials. As can be seen, rather small residual excitations varying with the transverse secular period are observed on the lower-frequency transverse mode, while the residual excitation is consistent with zero -irrespective of the dwell time- on the higher-frequency transverse mode.

\subsection{Separation and merging}\label{sec:separation}

\begin{figure*}
	\centering
	\includegraphics[width=0.7\textwidth]{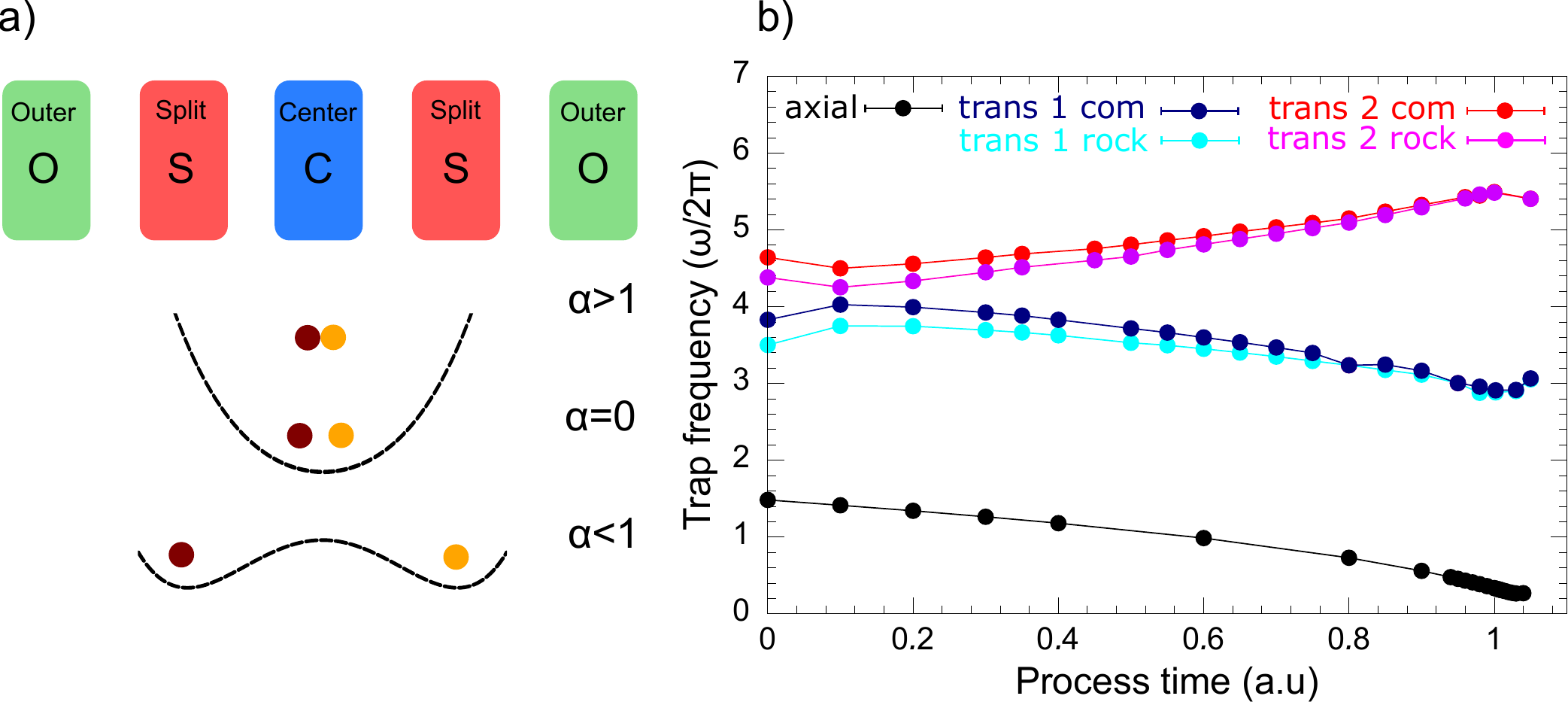}
	\caption{\textbf{a)} shows a cartoon of the separation process, where a single harmonic axial potential well is transformed into a double well. The involved electrodes are the outer electrodes $O$ serving for increasing the critical confinement, the split electrodes $S$ generating the double well and the center electrode $C$, where the ions are stored initially. \textbf{b)} shows how the secular frequencies vary up to slightly beyond the critical point. The axial confinement decreases down to about $2\pi\times\SI{200}{\kilo\hertz}$, while the transverse frequencies persist at values in the few-MHz range. It can also be seen how the transverse center-of-mass (com) and out-of-phase (rocking) modes become degenerate at the critical point, as the Coulomb coupling between the ions becomes weak at increasing distance. }
	\label{fig:trapfrequencyseparationcartoon}
\end{figure*}

Ion crystal separation constitutes the transfer of two ions stored in one common axial potential well into two separate individual wells at distinct trap segments, and merging is the reverse process. These operations require the transform of a harmonic potential well into a double-well potential, which necessarily involves a transient situation where the ions experience a strongly reduced axial confinement, see Fig. \ref{fig:trapfrequencyseparationcartoon}. This makes it the most challenging shuttling operation, as the low confinement exposes the ions to increased heating and improperly controlled potentials. The process requires driving the ions through a quasi-singularity in terms of equilibrium distance versus time. As pointed out in \cite{Kaufmann2014}, this makes it impossible to attaining adiabatic shuttling conditions and leads to finite axial excitation even at slow process times. Separation operations have been demonstrated and analyzed in more detail in Refs. \cite{Rowe2002,Bowler2012,Ruster2014}. \\

\begin{figure}
	\centering
	\includegraphics[width=0.45\textwidth]{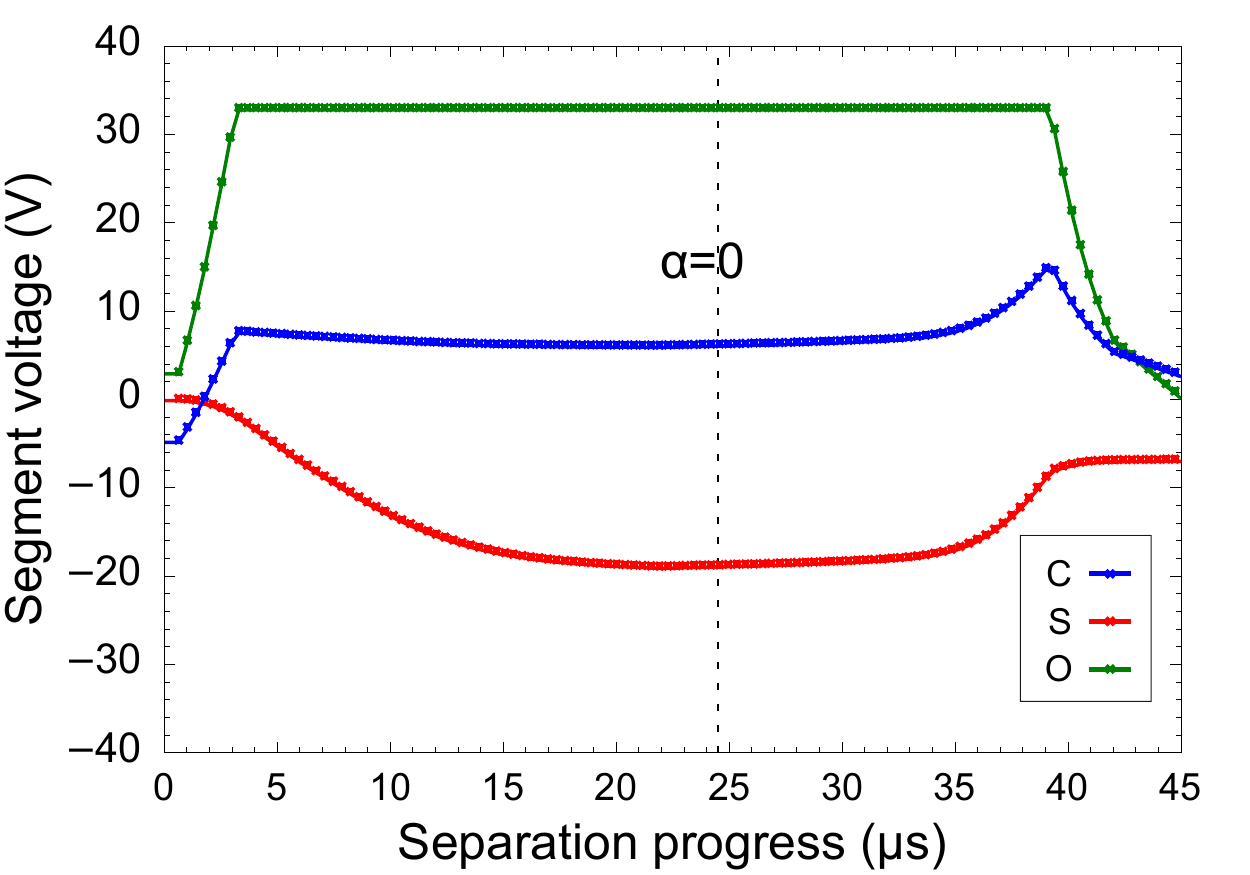}
	\caption{Separation voltage ramps. The blue/red/green curves correspond to the voltages applied to the C/S/O segments. The first part consists of a rapid decrease of the harmonic confinement. The middle part, featuring rather slow variation of the voltages, corresponds to slow driving of the ion crystal through the critical point. In the final part, the seperated ions are shuttled to the destination positions at the $S$ electrodes, which basically corresponds to simple linear transport.}
	\label{fig:separationramp}
\end{figure}

The residual confinement at the critical point, where the harmonic confinement at the symmetry center of the axial electrostatic potential vanishes, is caused by the fact that the Coulomb repulsion pushes the ions away from it, where the anharmonicity of the emerging double-well leads to nonzero potential curvature. Therefore, small trap geometries with large anharmonicity parameters are beneficial for conducting such operations.
For optimum excitation values, it is crucial to drive the ions slowly through the critical point, which in turn requires precise control over the harmonic confinement generated by the involved control electrodes. To that end, a precise calibration of the dependence of the axial secular frequency on the control voltages is obtained from resolved sideband spectroscopy. This way, tolerable axial excitations in the range of 1-10 phonons per ion and per operation are achieved. However, the spectroscopic calibration requires optical access to the site where the separation/merge operations take place, therefore these operations are so far restricted to the LIZ. Suitable control voltage ramps are shown in Fig. \ref{fig:separationramp}.\\

\begin{figure*}
	\centering
	\includegraphics[width=0.7\textwidth]{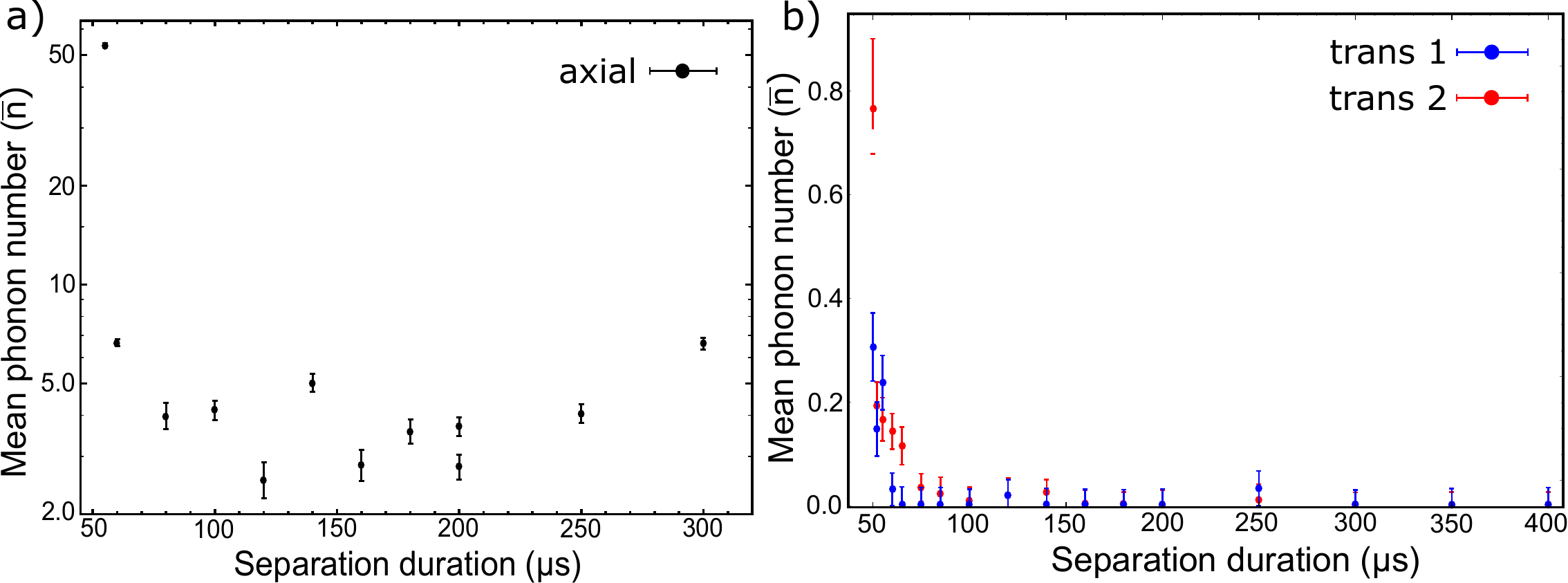}
	\caption{Separation induced excitation: \textbf{a)} shows the mean axial excitation per ion in terms of mean phonon number after separation. It can be seen that excitations in the few-phonon regime cannot be avoided. \textbf{b)} shows the transverse excitations per ion in terms of mean phonon number after separation on both modes. For sufficiently slow separation, the separation-induced excitation drops below the residual thermal excitation after sideband cooling.}
	\label{fig:separationexcitation}
\end{figure*}

A detailed account on how to obtain voltage ramps for good separation/merge performance and a characterization of the resulting axial excitation can be found in \cite{Ruster2014}. In this work, we present more recent results from an improved apparatus and an additional characterization of the transverse excitation induced by this process.\\
The resulting axial excitation versus process time is displayed in Fig. \ref{fig:separationexcitation} a). Strong excitation towards shorter process times can be observed, and for long times the excitation saturates due to the fact that the ions spend more time at low confinement and therefore are prone to increased anomalous heating. The more relevant parameter is the resulting transverse excitation, shown in Fig. \ref{fig:separationexcitation} b), as it is affecting the entangling gate fidelities. For sufficiently long times beyond $\SI{70}{\micro\second}$, no significant excitation is observed on both transverse modes of the singled ions, i.e. the measured values are consistent with the baseline values after sideband cooling. For shorter times, also the transverse excitations show a strong increase, and the process is ultimately failing for times shorter than $\SI{50}{\micro\second}$ due to ion loss. The presumed reason is that the voltage waveforms are distorted by the low-pass filters, see Sec. \ref{sec:filtersandwiring}. Faster separation/merging can thus eventually be achieved by compensating the filter-induced distortions and/or increasing the filter cutoff while simultaneously improving the noise characteristics of the waveform generator.

As pointed out in \cite{Ruster2014}, separation/merge processes are extremely sensitive to uncompensated axial background forces, which can be of electrostatic or ponderomotive nature. A suitable axial compensation field can be generated by applying a differential voltage to the $S$ or $O$ segments, see Fig. \ref{fig:trapfrequencyseparationcartoon} a). Its value is determined by determining the window of successful separation versus the differential 'tilt' voltage, and choosing the center of this window as the value yielding the best results. In the context of multi-qubit operations, care has to be taken to not generate undesired background fields from additional wells storing spectator ions. Possible solutions for this are to either determine the tilt voltage for each separation/merge operation in the sequence, or to conduct these operations with spectator wells created at fixed positions. The tilt voltage also serves as a control parameter for separation/merging operations on larger ion crystals. In Ref. \cite{Bowler2012}, it has been demonstrated that $M$ ions can be separated from an $N$ ion crystal, in a controlled manner, via tuning of the tilt voltage.

\subsection{Swapping}\label{sec:swapping}
Physical rotation of ion crystals, termed ion swapping, enables reordering of the qubit arrangement along the trap axis. This has first been demonstrated in \cite{Splatt2009} and used in conjunction with quantum logic operations in \cite{Kaufmann2017swap}. The underlying mechanism is as follows: Upon increasing the axial secular frequency above the lower transverse one, a two-ion crystal will undergo a structural transition to vertical alignment, along the direction of the weaker transverse confinement. Transient application of a diagonal potential via the neighboring electrodes tilts the ion crystal before and after crossing the structural transition, and therefore turns this process into a controlled rotation. For identical qubit ions, the physical rotation is equivalent to a logical SWAP gate. Therefore, the capability to perform fast swap operations can improve the effective connectivity of our 1D architecture, which has a beneficial impact on the overall resource requirements for conducting a given quantum algorithm. Furthermore, swapping can mitigate the requirement of shuttling ions through trap junctions, and facilitate the operation of multi-species registers.\\

\begin{figure*}
	\centering
	\includegraphics[width=0.6\textwidth]{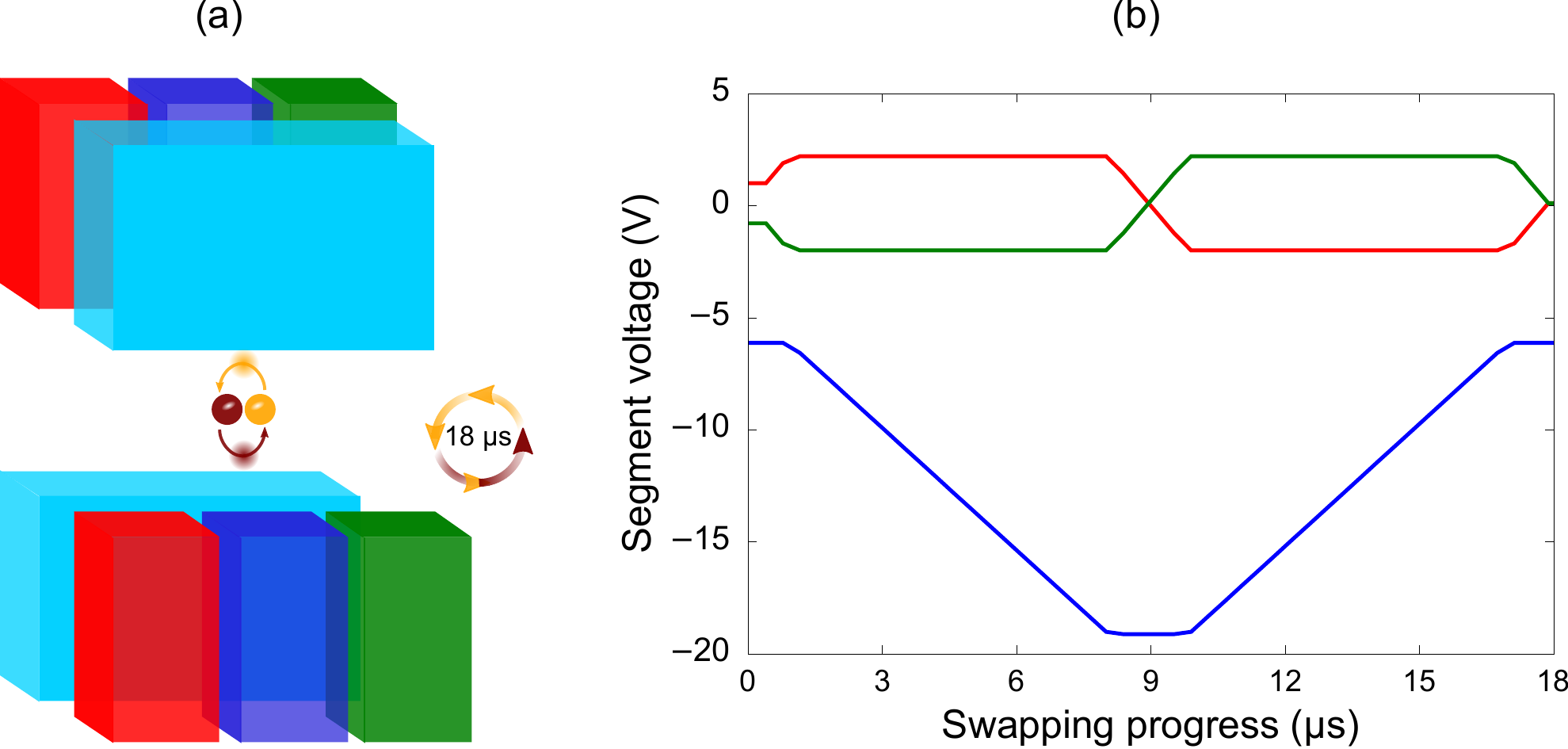}
	\caption{Ion crystal rotation. \textbf{a)} shows the trap electrodes controlling the rotation process: The center segment (blue) providing the axial confinement, the electrodes generating the tilt field (red/green) and the rf electrodes providing transverse confinement (light blue). \textbf{b)} shows the applied voltage ramps: First, the tilt field is ramped up by applying voltages of opposite polarity to the tilt electrodes. Then, the axial confinement is increased by increasing the negative bias on the center segment. After crossing the structural transition to vertical alignment of the ion crystal, the polarity of the tilt is reversed, and the crystal is restored to horizontal alignment. }
	\label{fig:swappingramp1}
\end{figure*}

Unlike the linear transport and separation/merge operations, swapping necessarily involves motion along the transverse directions, and therefore leads to transverse excitation even in the ideal case without technical imperfections. However, in contrast to the separation/merge processes, it is possible to maintain high secular frequencies along all directions throughout the entire process, therefore it can be performed fast and at tolerable transverse excitations. The resulting coherent oscillatory excitation in the ideal case will predominantly affect the axial out-of-phase mode, originating from the centrifugal force, and the transverse out-of-phase mode pertaining the weaker confinement direction, originating directly from the rotation movement. We verify low motional excitation by performing a single rotation on a two-ion crystal, which is cooled close to the ground state on all six secular modes before the rotation. Determining mean phonon number from fitting Rabi oscillations is more complicated in the two-ion case, therefore the simplified scheme described in Sec. \ref{sec:thermometrysimple} has been employed. The resulting values for the motional excitation on all six secular modes of a two-ion crystal after swapping are shown in Table \ref{tab:swapexcitation}. These values pertain to a swap duration of $\SI{18}{\micro\second}$.\\

\begin{table*}
\centering

\begin{tabular}{|C{3cm}|C{4cm}|C{2cm}|C{2cm}|C{2cm}|}
		\hline
		Secular mode & Secular frequency$/2\pi$ (MHz) &\textit{Lamb-Dicke ($\eta$)} & $\bar{n}$ after swapping & increase \\
		\hline
		axial:com&1.49&0.16(1)&0.36(2)&0.10(3) \\
		axial:str&2.58&0.11(1)&0.28(2) & 0.08(2) \\
		\hline
		trans1:com&3.81&0.06(1)&0.34(2)&0.01(3)\\ 
		trans1:rocking&3.49&0.06(1)&0.37(2)&0.12(3)\\ 
		\hline
		trans2:com&4.63&0.065(3)&0.25(2)&0.04(2)\\ 
		trans2:rocking&4.37&0.065(3)&0.23(2)&0.01(2)\\ 
		\hline
		\hline
\end{tabular}

\caption{Motional excitation in mean phonon number induced by a single rotation operation on a two-ion crystal. The values are obtained from averaging red sideband excitation with respect to probe time, see Sec. \ref{sec:thermometrysimple}.}
\label{tab:swapexcitation}
\end{table*}

\begin{figure*}
\centering
\includegraphics[width=0.7\textwidth]{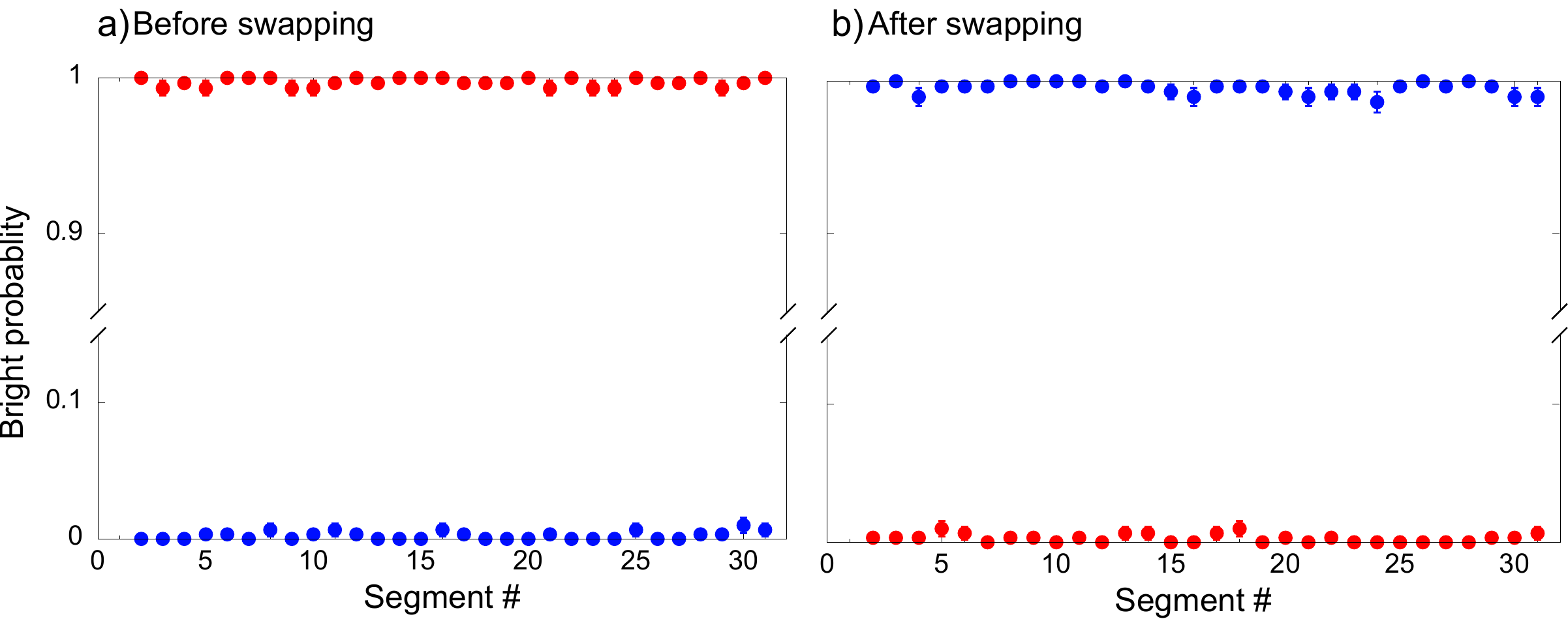}
	\caption{Remote swapping: a swap operation on a two ion crystal is conducted at different segments, and is verified by measuring the spin configuration. Blue (red) dots correspond to the result for ion 1 (2). (a) shows the fluorescence result without swap, and (b) shows the fluorescence result with swap, where 'bright'('dark') corresponds to $\ket{\downarrow}$($\ket{\uparrow}$). State preparation and measurement (SPAM) errors were characterized in \cite{Kaufmann2017swap} to be 0.4$\%$.}
	\label{fig:remoteswap}
\end{figure*}

The arbitrary waveform generator described in Sec. \ref{sec:bertha} allows for negative bias voltages of up to $\SI{-40}{\volt}$ at the center segment, therefore it is possible to conduct this process while maintaining high transverse secular frequencies in the few-MHz range. The fully independent control of all trap electrodes allows for swap operations at any trap site, and even for conducting multiple swap operations simultaneously. As all secular trap frequencies remain high during the process, a precise calibration as in the case of separate / merge operations is not required. Therefore, the process is not restricted to the LIZ. We have shown swapping of a two-ion crystal across the entire trap, the results are displayed in Fig.~\ref{fig:remoteswap}. Here, a two-ion crystal is separated, and the ions are consecutively moved to the LIZ, where ion $1$ is prepared in $\ket{\uparrow}$ and ion $2$ in $\ket{\downarrow}$. The ions are merged again and jointly shuttled to a given destination segment. There, the swap operation is performed. Then, the crystal is again separated at the LIZ, and both ions are consecutively moved to the LIZ for spin readout. It can be seen that, within the given readout fidelity, the spin configuration is indeed reversed for all destination segments, if the swap operation was actually carried out.

\begin{figure*}
	\centering
	\includegraphics[width=0.7\textwidth]{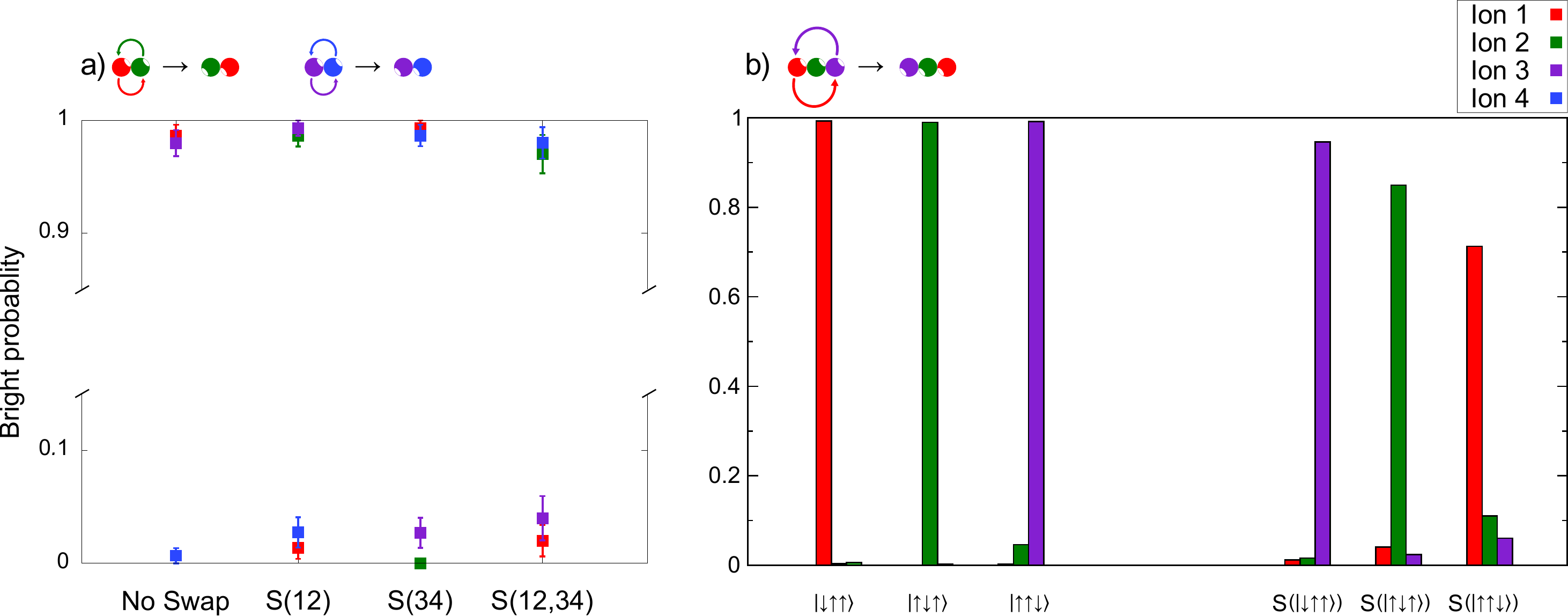}
	\caption{Simultaneous  three-ion crystal swapping:(a) shows the fluorescence results for no swapping, swapping of only ions $1,2$, swapping of only ions $3,4$ and swapping of $1,2$ and $3,4$ of a four ion register initialized to $\ket{\downarrow_1\uparrow_2\downarrow_3\uparrow_4}$. (b) shows the results for rotation of a three-ion crystal initialized in three different spin configurations.}
	\label{fig:swap_parallel_and_3}
\end{figure*}

Based on these results, we also demonstrate simultaneous swapping: Here, a four-qubit register is prepared in the spin configuration $\ket{\downarrow_1\uparrow_2\downarrow_3\uparrow_4}$. The ions are then separated into two-ion crystals consisting of ions $1,2$ and $3,4$. These pairs are moved to distinct segments at a distance of at least three segments. Measurements with different reconfiguration operations are performed: Ion swapping is attempted either on crystal one, crystal two or simultaneously on both. For comparison, a measurement is also performed with no ion swapping. Then, in each case, the ion crystals are separated into individual ions, which are consecutively read out.  The results are shown in Fig. \ref{fig:swap_parallel_and_3} a). It can again be seen that the spin configuration is deterministically permuted based on which swap operations are actually driven, with the fluctuations being consistent with readout errors.

We also show that reordering via crystal rotation can be extended to larger crystals. In Fig. \ref{fig:swap_parallel_and_3} b), we show results for rotation of a three-ion crystal. Similar to the protocols above, the crystal is initialized to the three different spin configurations $\ket{\downarrow_1\uparrow_2\uparrow_3}$, $\ket{\uparrow_1\downarrow_2\uparrow_3}$ or $\ket{\uparrow_1\uparrow_2\downarrow_3}$. The spin is read out in the original arrangement sequence, for the cases that the rotation is attempted or not. For the case with operation carried out, it can be seen that the spin population is predominantly in the expected state, i.e. the spin state of ions $1$ and $3$ is swapped. However, in this case the errors exceed the readout errors, presumably as increased transverse excitation deteriorates the spin readout fidelity. The three-ion rotation has been carried out using the same voltage ramps as for the two-ion rotation. Due to the increased axial extent of the three-ion crystal, the forces acting throughout the rotation are correspondingly increased. However, with a dedicated optimization of the ramps, it should be possible to carry out high-fidelity rotation operations on register building blocks comprised of three or more ions.

\section{Outlook}\label{sec:outlook}
In this review, we have given an overview of scalable trapped-ion quantum computer architectures based on shuttling operations. We have discussed the relevant technological components required to construct and operate such a node, in particular segmented microchip ion traps and fast multichannel arbitrary waveform generators. We have described the design, characterization and interplay of the relevant components and operations in detail, outlining how a fully functional few-qubit quantum processing node based on ion shuttling is accomplished. In order to further boost the performance of such a platform, a number of well-defined technological and methodological building blocks is to be implemented. Most of these techniques have already been successfully demonstrated by different ion trapping groups, however mostly in a separate experiments.

Further improvements of the control hardware are required to allow for real-time conditional branching of computational sequences, based on the results of in-sequence measurements. A suitable execution control system will be able to conduct arbitrary branched sequences composed of shuttling and gate operations. Implementing such a system will make the node capable of autonomously performing a large number of error-correction cycles, rendering it to be fault tolerant. Other measures have to be implemented to improve the operational fidelities further, ranging from increased laser powers and composite pulse sequences to improve gate fidelities, to improvements of the qubit coherence times e.g by using hyperfine qubits, e.g. $^{43}$Ca$^+$ \cite{Harty2014}, together with ultra-stable magnetic fields \cite{Ruster2016}. 

Adding at least one additional ion species will enable sympathetic cooling  and selective readout of subsets of the qubit register. Both schemes are based on the fact that the auxiliary ion species allows for resonant driving with light which is far off-resonant to any relevant optical dipole transition of the qubit species. This has for example been successfully demonstrated by the groups at NIST, in Oxford and in Zurich \cite{Tan2015, Ballance2015, Negnevitsky2018}. However, adding an auxiliary species, which was demonstrated in \cite{Home2009} will lead to an 
increase in controlling the shuttling operations.

Finally, to realize a scalable trapped-ion quantum computer capable of outperforming classical computers with useful tasks, requires an array of fully automated 
processing nodes, each consisting of one ion trap. Optical interfaces are a possible means for interconnecting the nodes within such an array, and could be based on free-space or cavity-enhanced coupling to optical radiation. Several research groups have realized free-space \cite{Streed2011, Hucul2014, Stephenson2019} optical interfaces, and work on cavity-based optical interfaces is ongoing \cite{Brandstaetter2013, Pfister2016, Takahashi2018}. Another approach could be to combine segmented linear traps with the extraction of single ions \cite{Jacob2016} from one trap and its injection \cite{Groot2019} into a second ion trap. This scheme would be similar to a shuttling based approach \cite{Lekitsch2017}, however based on trap of modest complexity. In conclusion, shuttling-based trapped-ion QC has been shown to be a suitable approach to noisy intermediate-scale quantum computing. Methodological building blocks and the overall working principle have been demonstrated successfully, and further improvements based on existing technology could be accomplished in the foreseeable future. 

While the steps outlined above and the integration into one platform are challenging, an optically connected network of ~10 quantum processing nodes each bearing 10-100 trapped-ion qubits is within reach of today's technology. Such an apparatus could already allow for outperforming classical computing hardware on useful tasks such as computing molecular energy structures \cite{Hempel2018}. 

As the operational timescales of such a platform will be dominated by the shuttling-induced overhead, which is in turn tied to the timescales imposed by the secular trap frequencies, such an architecture will not be competitive to e.g. superconducting platforms in terms of operational clock rate. However, the effective speed is also determined by the operation fidelities and the effective connectivity, moreover the performance advantage of a quantum computer is given by its intrinsic \textit{quantum parallelism}. Therefore, it is promising to keep on pursuing the shuttling-based trapped-ion approach to scalable quantum computing.

\begin{acknowledgments}
The research is based upon work supported by the Office of the Director of National Intelligence (ODNI), Intelligence Advanced Research Projects Activity (IARPA), via the U.S. Army Research Office grant W911NF-16-1-0070. The views and conclusions contained herein are those of the authors and should not be interpreted as necessarily representing the official policies or endorsements, either expressed or implied, of the ODNI, IARPA, or the U.S. Government. The U.S. Government is authorized to reproduce and distribute reprints for Governmental purposes notwithstanding any copyright annotation thereon. Any opinions, findings, and conclusions or recommendations expressed in this material are those of the author(s) and do not necessarily reflect the view of the U.S. Army Research Office. We acknowledge funding from the EU H2020-FETFLAG-2018-03 under Grant Agreement no.820495 and by the Germany ministry of science and education (BMBF) via the VDI within the project VERTICONS.
\end{acknowledgments}

\bibliographystyle{apsrev4-1}
\bibliography{STIQP}
\end{document}